\documentclass{aa}
\usepackage{graphicx}
\usepackage{natbib}
\bibpunct{(}{)}{;}{a}{}{,}
\usepackage{graphicx}
\usepackage{revsymb}	
\usepackage{amsmath}	
\usepackage[usenames]{color}	
\usepackage{epstopdf}	
\newcommand{\inv} {\frac {1}}

\newcommand{\deriv} [2] {\frac {\textrm{d} #1 } {\textrm{d} #2} }

\newcommand{\eq}[1] {Eq.\,(\ref{#1})}
\newcommand{\eqn} [1] {
\begin{equation} #1
\end{equation}}

\begin{document}
\title{
Stochastic excitation of nonradial modes\\
II. Are solar asymptotic gravity modes detectable?
}
   \author{}
   \author{K.~Belkacem \inst{1}
           \and
           R.~Samadi \inst{1}
            \and
           M.~J.~Goupil \inst{1}
           \and
           M.~A.~Dupret\inst{1}
           \and
           A.~S.~Brun \inst{2,3}
           \and
           F.~Baudin\inst{4}
          }

 \institute{Observatoire de Paris, LESIA, CNRS UMR 8109, 92195 Meudon, France 
 \and
DSM/IRFU/SAp, CEA Saclay \& AIM, UMR 7158, CEA - CNRS - Universit\'e Paris 7, 91191, Gif-sur-Yvette, France
\and
 Observatoire de Paris, LUTH, CNRS UMR 8102, 92195 Meudon, France
 \and
  Institut d'Astrophysique Spatiale, CNRS - Universit\'e Paris XI UMR 8617,91405 Orsay Cedex, France}

   \offprints{K. Belkacem}
   \mail{Kevin.Belkacem@obspm.fr}
   \date{\today}

  \authorrunning{Belkacem et al.}
  \titlerunning{Stochastic excitation of nonradial modes II. Are solar asymtotic gravity modes detectable? }

  \abstract
  {Detection of solar gravity modes 
   remains  a major challenge to our understanding of the inner parts of the Sun. Their frequencies would enable the derivation          
   of constraints on the core physical properties, while their amplitudes can put severe constraints on the properties 
   of the inner convective region.}
  {Our purpose is to determine accurate theoretical  amplitudes of solar g modes  and estimate the SOHO 
  observation duration  for an unambiguous detection of individual modes. We also explain  
  differences in theoretical amplitudes derived from previous works.}
  { We investigate the stochastic excitation of modes by turbulent convection, as well as their damping. 
  Input from a 3D global simulation of the solar convective zone is used for the kinetic turbulent energy spectrum. 
  Damping is computed using a  parametric description of the  nonlocal, time-dependent, convection-pulsation interaction.  
   We then provide a theoretical estimation of the intrinsic, as well as  apparent, surface velocity.} 
  {Asymptotic g-mode  velocity amplitudes are found  to be orders of magnitude higher than previous works.  
    Using a 3D numerical simulation from the ASH code, we attribute this  
    to the temporal-correlation between the modes and the turbulent
    eddies, which  is found to follow a Lorentzian law rather than a Gaussian one, as previously used. 
    We also find that damping rates of asymptotic gravity modes are dominated by radiative losses, with a
    typical life time of $3 \times 10^5$ years for the $\ell=1$ mode at 
    $\nu=60\,\mu$Hz. The maximum velocity in the considered frequency range (10-100 $\mu$Hz) is obtained for the $\ell=1$
     mode at $\nu=60\,\mu$Hz and for the $\ell=2$ 
     at $\nu=100 \,\mu$Hz.
     Due to uncertainties in the modeling, amplitudes at maximum i.e. for
   $\ell=1$ at 60 $\mu$Hz can range from 3 to 6 mm s$^{-1}$. The upper limit is too high, as g modes would have been easily detected
    with SOHO, the GOLF instrument, and this sets an upper constraint mainly on  the convective velocity in the Sun.} 
  {}

   \maketitle
%
%________________________________________________________________

\section{Introduction}
\label{intro}

The pioneer works of \cite{Ulrich70} and \cite{Leibacher71} led to the identification of
 the solar five-minutes oscillations as global acoustic standing waves ($p$~modes). 
Since then, successful works have determined the Sun internal structure from the
knowledge of its oscillation frequencies \citep[e.g., ][]{CD04}. 
However, $p$ modes are not well-suited to probing the deepest  inner part of the Sun. On the other hand, 
 $g$ modes are mainly trapped in the radiative region and are thus  
able to provide information on the properties of the central part of the Sun ($r < 0.3 \, R_\odot$) 
\citep[e.g., ][]{TC01,CD06}. As $g$ modes are evanescent in the convective region,
their amplitudes are expected to be very low at the photosphere and above, where observations are made, 
making their detection is thus quite a challenge for more than $30$ years. 

The first claims of detection of solar gravity modes began with the work of \cite{Severny76} and 
\cite{Brookes76}. Even after more than ten years of observations from SOHO, 
there is still no consensus about detection of solar $g$ modes.
Most of the observational efforts have been focused on low-order $g$ modes motivated by a low 
the granulation noise \citep{Appourchaux06,Els06} and by previous theoretical 
estimates of $g$-mode amplitudes \citep[e.g.,][]{TC04,Kumar96}. 
Recently, \cite{Garcia07} have investigated the low-frequency domain, with the hope of 
detecting high radial-order $g$ modes. The method looked for 
regularities in the power spectrum, and the authors  claim to detect a periodicity 
in accordance with what is expected from  simulated power spectra. 
The work of \cite{Garcia07}  present the advantage of exploring 
a different frequency domain ($\nu \in [25;140] \mu$Hz) more favorable to a reliable theoretical 
estimation of the $g$-mode amplitudes, as we will explain later on.

Amplitudes of $g$ modes, as $p$~modes,  are believed to result from a balance between driving and 
damping processes in the solar convection zone. 
Two major processes have been identified as stochastically  driving the resonant 
modes in the stellar cavity. The first is related to the Reynolds stress tensor,  the second
is caused by the advection of turbulent fluctuations of entropy by
turbulent motions.
Theoretical estimations based on stochastic excitation have been previously 
obtained  by \cite{Gough85} and \cite{Kumar96}. 
\cite{Gough85} made  an order of magnitude estimate based on an assumption of  equipartition 
of energy as proposed by \cite{GK77}. He found a maximum 
velocity around $0.5 \, \textrm{mm\,s}^{-1}$ for  an $\ell=1$ mode at $\nu \approx 100\, \mu$Hz. 
\cite{Kumar96} used a different 
approach based on the \cite{GK94} modeling of stochastic excitation by turbulent convection,  
as well as an estimating of the damping rates \citep{Goldreich91} that led to a 
surface velocity near $0.01\, \textrm{mm\,s}^{-1}$ for   the $\ell=1$ mode 
at $\nu \approx 100\, \mu$Hz. 
The results differ from each other by orders of magnitude, as pointed out by 
\cite{CD2002}. Such differences  remain  to be  understood. One purpose of the present work is to 
carry out  a  comprehensive study of both the excitation and damping rates of asymptotic 
$g$ modes.  Our second goal is to provide theoretical oscillation mode velocities, , as reliably as possible.
Note, however, that penetrative convection is another possible excitation mechanism \citep{Andersen96,Boris05}, 
but it is beyond the scope of this paper.  

Damping rates are computed using 
the \cite{MAD05} formalism that is based on a non-local time-dependent 
treatment of convection. We will show that, contrary to $p$~modes and high frequency 
$g$~modes, asymptotic $g$-mode (i.e low frequency) damping rates are insensitive 
to the treatment  of convection. This then removes most of the uncertainties in the estimated 
theoretical oscillation mode  velocities.  
Consequently, we restrict our investigation to low-frequency gravity modes. 
Stochastic excitation is modeled in the same way as in   
\cite{Belkacem08}, which is a generalization to non-radial modes 
 of the formalism  developed by \cite{Samadi00I} and \cite{Samadi02I,Samadi02II}, for radial modes.  
As in the case of  $p$-modes,  the excitation formalism requires knowing the 
  turbulent properties of the convection 
zone, but unlike  $p$ modes, 
the excitation of gravity modes is not concentrated  towards the uppermost
surface layers. One must then have some notion about the turbulent properties across the whole 
convection zone. Those properties will be inferred from 
a 3-D numerical simulation provided by the ASH code \citep{Miesh08}.

The paper is organized as follows. Section~2 briefly recalls our model 
for the excitation by turbulent convection and describes the input 
from a 3D numerical simulation. Section~3 explains how 
the damping rates are computed. 
Section~4  gives our theoretical  results on the surface velocities of asymptotic  $g$~modes 
and compares them  with those from previous works.
Section~5 provides the apparent surface velocities, which take disk integrated effects 
and line formation height into account. These quantities can be directly compared with observations.
  We then discuss our ability to detect 
these modes  using data from the GOLF instrument onboard SOHO as a function of  the observing duration.      
The discussion is based on estimations of detection threshold and numerical simulations  of power spectra. 
In Sect.~6., uncertainties on the estimated theoretical and apparent velocities, due to the main uncertainties in our
modeling,  are discussed. Finally, conclusions 
are provided  in Sect.~7.

\section{Excitation by turbulent convection}

%\subsection{Theoretical formalism}

The formalism we used to compute excitation rates of non-radial modes 
was developed by \cite{Belkacem08} who   
extended the work of \cite{Samadi00I} 
developed for radial modes to non-radial modes. 
It takes the two sources into account that drive the resonant modes of the stellar 
cavity. The first is related to the Reynolds stress tensor and the second 
one is caused by the advection of the turbulent fluctuations of entropy by
the turbulent motions (the ''entropy source term''). 
Unlike for  $p$~modes, the entropy source term is negligible for $g$ modes. 
We numerically verified that it  
is two to four orders of magnitude lower than the 
Reynolds stress contribution depending on frequency.
This is explained by the entropy contribution being sensitive to  
second-order derivatives of the displacement eigenfunctions in the superadiabatic region where 
entropy fluctuations are localized. As the gravity modes 
are evanescent in the convection zone, the second derivatives 
of displacement eigenfunctions are negligible and  so is  the entropy 
contribution.

The excitation rate, $P$,  then  arises from the   Reynolds stresses  and can 
be written  as \citep[see Eq.~(21) of][]{Belkacem08} 
\begin{align}
\label{puissance}
P &= \frac{\pi^{3}}{2  I} \int_{0}^{M}  \textrm{d}m   \, \rho_0  \, R(r) \int_{0}^{+\infty}  
\textrm{d}k \; \mathcal{S}_k \\
\mathcal{S}_k &= \frac{1}{k^2 } \int_{-\infty}^{+\infty} \textrm{d}\omega~E^2(k)  
~\chi_k( \omega + \omega_0) ~\chi_k( \omega )
\end{align}
where $m$ is the local mass, $\rho_0$ the mean density, $\omega_0$ the mode angular 
frequency, $I$ the mode inertia,
$\mathcal{S}_k$ the source function,  $E(k)$ the spatial kinetic energy 
spectrum, $\chi_k$ the eddy-time correlation function, and $k$ the wavenumber. 
The term $R(r)$ depends on the eigenfunction, its expression is given in Eq.~(23) of \cite{Belkacem08}, 
i.e 
\begin{eqnarray}
\label{gammabeta}
R (r) &=&   {16\over 15} ~  \left| \deriv{\xi_r}{r}  \right|^2  +  {44\over 15} ~    \left| \frac{\xi_r}{r}  \right|^2
+   \frac{4}{5} \left( \frac{\xi^*_r}{r} \deriv{\xi_r}{r} + c.c \right) \nonumber \\ 
&+&  ~ L^2 \left( {11\over 15} ~  \left| \zeta_r  \right|^2 - {22 \over 15} (\frac{\xi_r^* \xi_h}{r^2} +c.c) \right) \nonumber \\
&-&{2\over 5} L^2 \left(\deriv{\xi^*_r}{r} {\xi_h \over r} 
+ c.c \right)\nonumber \\
&+&  \left| \frac{\xi_h}{r}  \right|^2 
\left( \frac{16}{15} L^4+ \frac{8}{5} {\cal F}_{\ell,\vert m \vert} - \frac{2}{3} L^2 \right) \, ,
\end{eqnarray}
where we have defined
\begin{eqnarray}
L^2 &=& \ell (\ell + 1) \\
\zeta_r &\equiv&  \deriv{\xi_h}{r} +\frac{1}{r}(\xi_r-\xi_h) \\
{\cal F}_{\ell,|m|} &=&  \frac{|m| (2 \ell + 1)}{2} \left( L^2- (m^2 + 1 )\right) \, ,
\end{eqnarray}
and $\xi_r, \xi_h$ are the radial and horizontal components of the fluid displacement eigenfunction ($\vec \xi$), and 
$\ell,m$ represent the degree and azimuthal number of the associated spherical harmonics.

\subsection{Numerical computation of theoretical excitation rates}
\label{compute_power}

%\subsubsection{Physical input}
%\label{compute}

In the following, we compute the excitation rates of $g$~modes  for a solar model.
The rate ($P$) at which energy is injected into a mode per unit time is
calculated according to \eq{puissance}. 
Eigenfrequencies and eigenfunctions  are computed
using the adiabatic pulsation code OSC \citep{boury75}.
The solar structure model used for these computations is
obtained with the stellar evolution code CESAM \citep{Morel97} for the
interior and a \cite{Kurucz93} model for the atmosphere.  The
interior-atmosphere matching point is chosen at $\log\tau=0.1$ (above the
convective envelope).  The pulsation computations use the full
model (interior+atmosphere).  In the interior model, we used the OPAL
opacities \citep{Opal96} extended to low temperatures with the
opacities of \cite{Alexander94} and the CEFF equation of state
\citep{CD1992}. Convection is included according  
to a B\"{o}hm-Vitense mixing-length (MLT) formalism
\citep[see][for details]{Samadi06}, from which the convective velocity 
is computed.  Turbulent pressure is not included (but see discussion in Sect.6).

Apart from the eigenfunctions and the density stratification, \eq{puissance} involves both the convective velocity and the 
turbulent kinetic energy  spectrum.
To get some insight into the turbulent properties of the {\it inner} part of the solar convection 
zone, we chose to use results from {\it (ASH)} 3D numerical simulations. 
Such a choice was motivated by the uncertainties inherent in the treatment of turbulence by 
the MLT. The MLT indeed only gives us an estimation of the convective flux but is not able 
to assess the contributions of all scales involved in turbulent convection. 
Thus, in the following, the rms convective velocity is taken from the mixing-length theory,  
while both the spatial and temporal turbulent properties are inferred from the 3D simulation. 
 Then, velocity from the numerical simulation is not used in our calculation. 
This choice is motivated by the rigid boundary condition at the top of the simulation that 
results in an unrealistic decrease in the vertical velocity for $r > 0.93 \, R_\odot$.

\subsection{The 3D convection simulation}
\label{sec:simu}
\begin{figure}[t]
\begin{center}
\includegraphics[height=6cm,width=9cm]{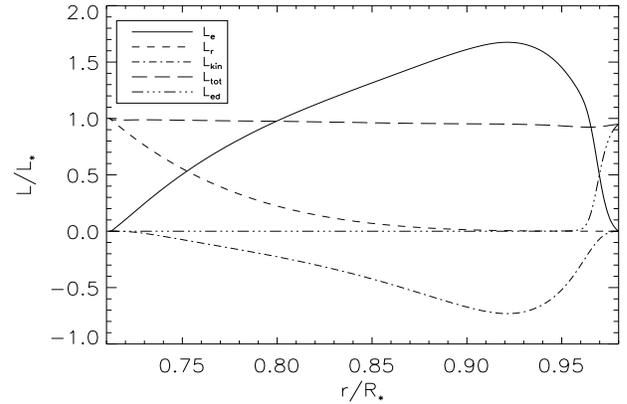}
\caption{Luminosity flux  contributions versus radius, averaged over horizontal surfaces and in time. The solid line 
corresponds to the enthalpy luminosity ($L_e$), the short dashed-line to the radiative luminosity ($L_r$), the dotted-dashed line 
to the kinetic energy luminosity ( $L_{kin}$), the long-dashed line to the total luminosity ($L_{tot}$), and the dot-dashed line correspond to the unresolved eddy luminosity ($L_{ed}$) \citep{Brun04}. We particularly emphasize the negative kinetic-energy flux that results in a larger convective flux (see text for details).}
\label{spectre}
\end{center}
\end{figure}

One way of assessing the dynamical properties of the deep solar turbulent convection zone is 
to exploit a high resolution numerical simulation such as those performed with the anelastic spherical harmonic
(ASH) code \citep{Miesh08,Brun04}. The simulation of global scale turbulent convection used in the present work 
 is discussed in detail in Miesch et al. (2008).
The ASH code solves the hydrodynamic anelastic equations within a spherical shell extending from $r=0.71$ up to $r=0.98 R_{\odot}$, yielding an overall radial density contrast of 132. Solar values were assumed for the rotation rate and the imposed luminosity.  
Figure 1 represents the energy flux balance (converted to luminosity and normalized to the solar luminosity) 
 in the simulation. We clearly see how dominant, and overluminous, the convective (enthalpy) flux 
is in carrying the heat outward. This is mostly due to the strong density contrast and to the corresponding
strong asymmetry between up- and downflows yielding a large inward kinetic energy flux (see Miesch et al. 2008 for more details).
We have seen above that, in order to compute the excitation rate of the waves,
one needs some well-defined physical quantities, such as the kinetic energy spectrum $(E_k)$ and the eddy time function $({\chi_k})$.
 It is straightforward to deduce these quantities from the 3D simulation as explained in Appendix B. 
 We then directly use $E_k$ in Eq. (2) to compute the source function, whereas for $\chi_k$ we perform 
 a fit of the 3-D results with a simple analytical expression.
In the ASH code, the set of anelastic equations is projected onto 
spherical harmonics for the horizontal dimensions. 
This implies that the kinetic energy spectrum is obtained as a function of the spherical 
degree {\cal \it l} . The local wavenumber $k_h$ is  obtained via 
the simple expression $k_h=\sqrt{{\cal \it l} ({\cal \it l} +1)}/r$, with $r$ the shell radius.

\subsubsection{Kinetic energy spectrum and time-correlation function}
\label{kin}

\begin{figure}[t]
\begin{center}
\includegraphics[height=6cm,width=9cm]{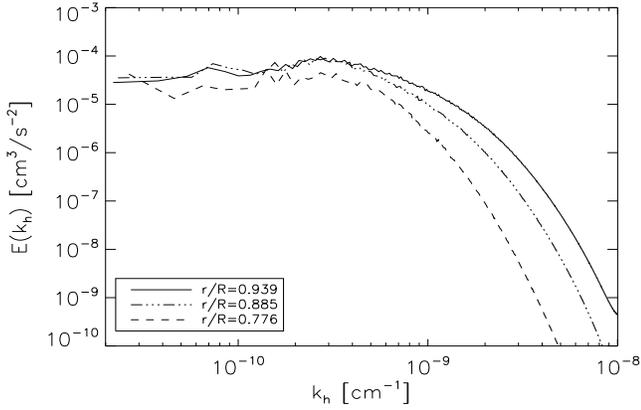}
\caption{$E(k_h)$ computed as explained in Appendix.~\ref{ASH}, for three shell radii that sample the convection zone, 
as a function of the local horizontal wave number $k_h$.}
\label{spectre}
\end{center}
\end{figure}

\begin{figure}[t]
\begin{center}
\includegraphics[height=6cm,width=9cm]{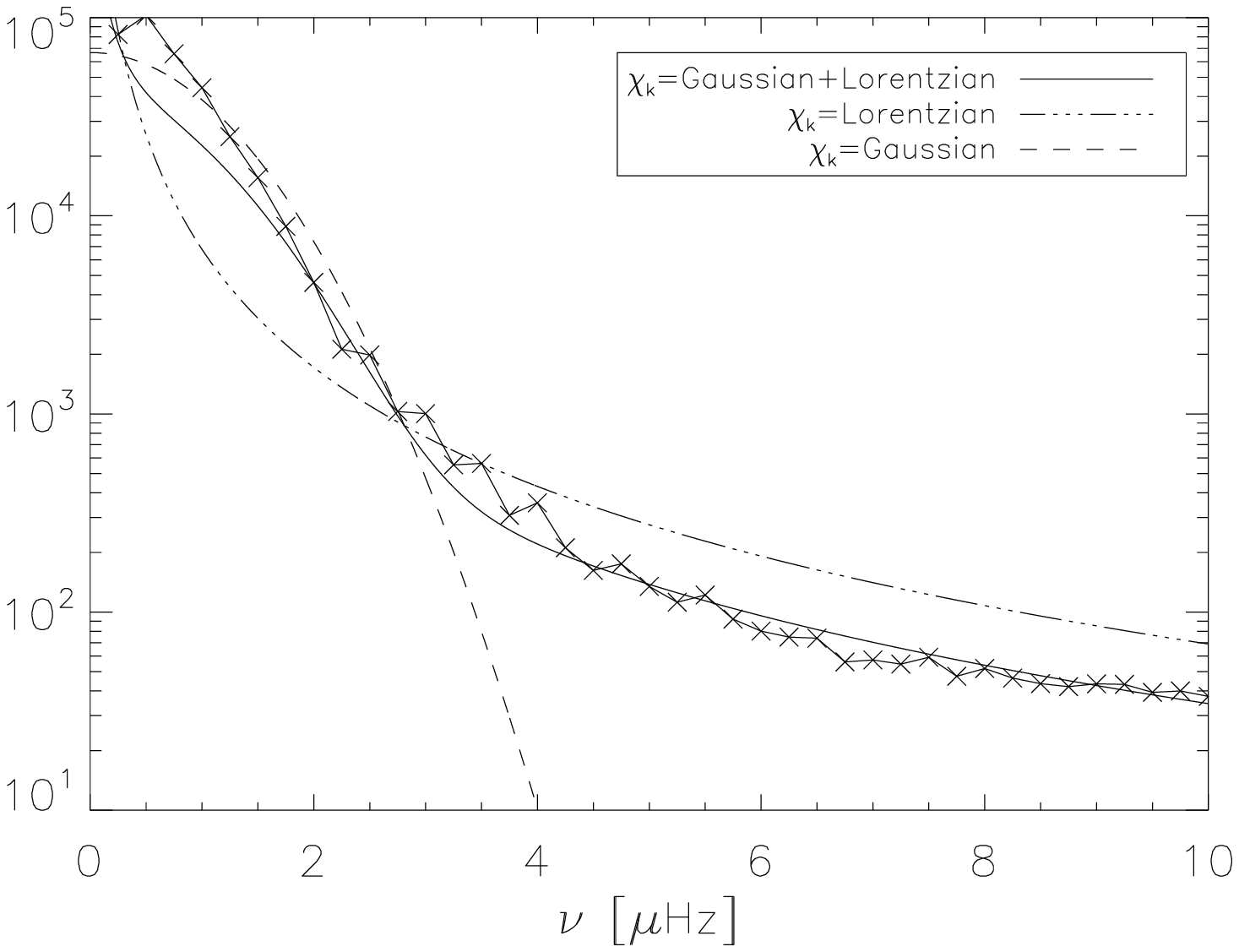}
\includegraphics[height=6cm,width=9cm]{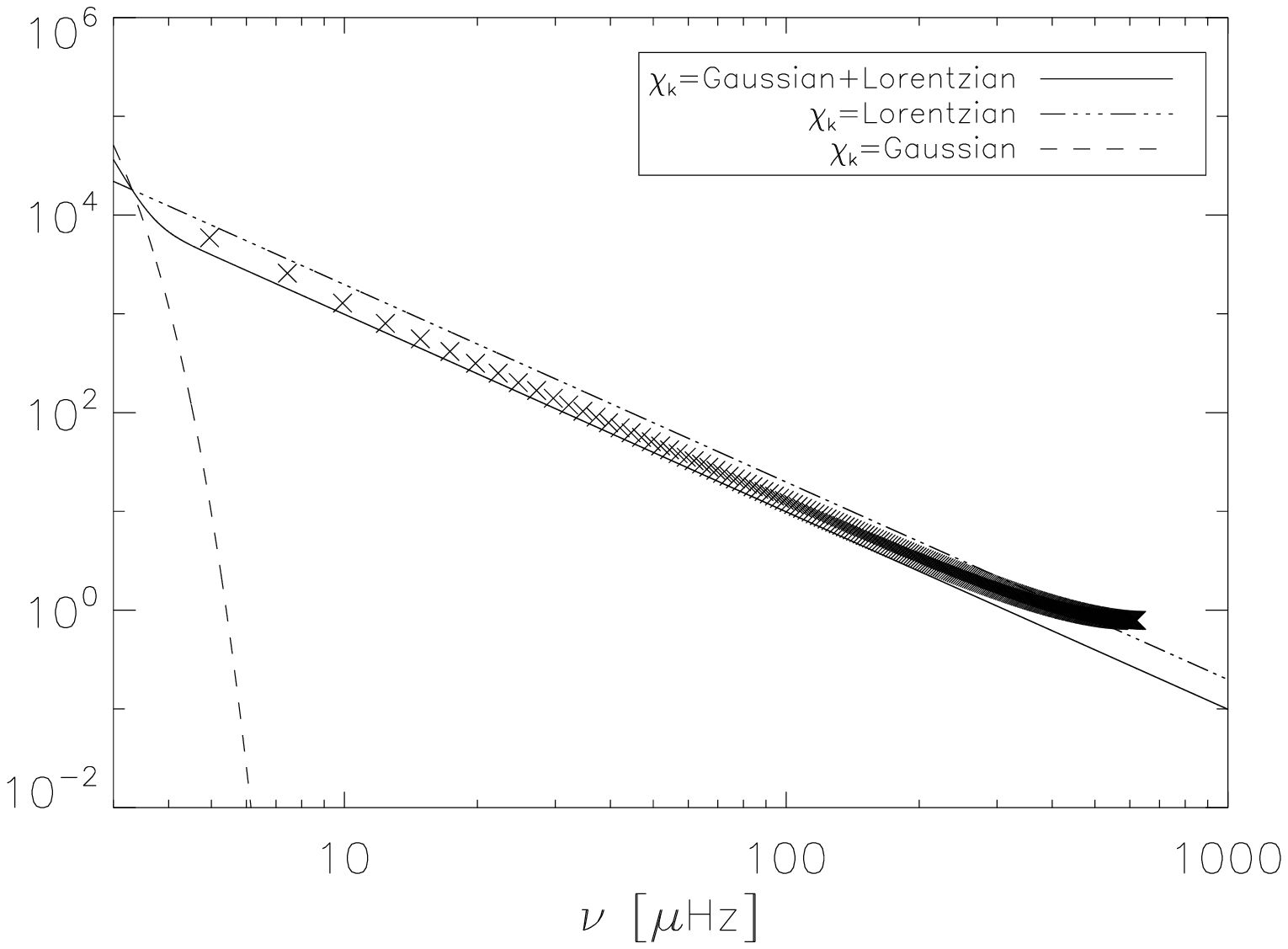}
\caption{{\bf Top:} Crosses represent $\chi_k(\omega)$ obtained from the 3D simulation 
at the wave number $k_0$ that corresponds to the maximum of $E(k)$, 
and at the radius $r/R_\odot = 0.89$. Data are obtained with a time series  of duration 
$\approx 45.83$  days with a sampling time of $4 \cdot 10^4$ seconds. 
Analytical functions are normalized so that their integrals are equal to unity. 
{\bf Bottom:}  The same as the upper panel except that data are obtained 
with a time series  of duration $\sim$ 4.68 days with a sampling time of 800 seconds. 
The theoretical curves are normalized so that their integrals over frequency equal
that of the simulated data.}
\label{chik_simu_vs_analytique}
\end{center}
\end{figure}

The kinetic energy spectrum of 
the total velocity ({\it i.e.} the horizontal and vertical components), $E(k_h)$,  is plotted
in the top panel of Fig.~\ref{spectre} as a function of the local horizontal wave number $k_h$. 
The rms convective velocity ($u$)
increases with  $r$, thus explaining that the deeper the layers,  the smaller 
$E(k)$ since $\int \textrm{d}k \, E(k) = 1/2 \, u^2$. In terms of excitation rates, an important issue is the scale at which the spectrum peaks. 
As pointed out by \cite{Miesh08}, the scale at which the  kinetic energy spectrum is maximum is  the scale between the  downwflows. 
It is about $58$ Mm at the top of the simulation ($r=0.98\,R_\odot$)  up to $300$ Mm at the bottom. 
This is quite different from what is found in the uppermost layers  
in 3D numerical simulations of the Sun \citep[e.g., ][]{Stein98}, 
in which the maximum of $E(k)$ is found on a scale around $1$ Mm. Such a difference is 
explained by the density that strongly decreases in the upper layers. 

The time-correlation function
($\chi_k$)  also plays an  important role. 
Usually, a Gaussian time-correlation function is used \citep{GK94,Chaplin05}. 
\cite{Samadi02II} demonstrate that $\chi_k$ is reproduced better by a Lorentzian function. They 
argue that the departure from a Gaussian function can be explained by the presence of plumes 
in the uppermost part of the convection zone. This result, obtained with 3-D numerical simulations, 
was then confirmed by confronting  solar-$p$ modes excitation rates, computed 
with Gaussian and Lorentzian functions, with the observational data. It turns out that the Lorentzian 
function greatly improves the agreement between models and  observations. 
However, the time-correlation function is unknown at deeper layers. 
The eddy-time correlation function derived from the 3D numerical simulation
 provided by the ASH code is therefore compared 
to Gaussian and Lorentzian functions that are respectively  defined as
\begin{align}
\chi_k (\omega ) &= \inv  { \omega_k \, \sqrt{\pi}}  e^{-(\omega / \omega_k)^2} \label{eqn:GF}\\
\chi_k (\omega ) &=\frac{1} {\pi \omega_k/2} \,\frac{1}{1+ \left( 2 \omega / \omega_k \right )^2}
\label{eqn:GF1}
\end{align}
with the condition
\begin{equation}
\int_{-\infty}^{+\infty} \chi_k (\omega ) \textrm{d}\omega = 1
\end{equation}
where $\omega_k$ is its linewidth, defined as
\eqn{
\omega_k  \equiv \,  {2 \,k \, u_k \over \lambda}  
\; .
\label{eqn:omegak}
} 
where $\lambda$ is a parameter as in \cite{B92c}, 
the velocity  $u_k$ of the eddy with wavenumber $k$ 
is related to the kinetic  energy spectrum $E(k_h)$ by   \citep{Stein67} 
\eqn{
u_k^2 =  \int_k^{2 k}  dk \, E(k) \; .
\label{eqn:uk2}
} 

Figure~\ref{chik_simu_vs_analytique} presents the comparison between analytical time-correlation 
functions, computed following the set of Eqs.~(\ref{eqn:GF})-(\ref{eqn:uk2}), 
and $\chi_k$ computed 
from the 3D numerical simulation. The latter is calculated as described in Appendix.~\ref{ASH}. 
The Lorentzian function represents the eddy-time correlation function better than 
a Gaussian function in the frequency range we are interested in  
($\nu \in [20 \, \mu\textrm{Hz}; 110 \, \mu\textrm{Hz}]$). 

The best fit is found using a sum of a Lorentzian function with $\lambda=3$ 
and a Gaussian with $\lambda=1/3$ as shown in the top panel of Fig.~\ref{chik_simu_vs_analytique}. 
In the frequency range we are interested in, i.e. at frequencies corresponding to 
the gravity modes (bottom panel of Fig.~\ref{chik_simu_vs_analytique}) the fit reproduces  
the time-correlation given by the 3-D numerical simulation. 
We also clearly see that the eddy-time correlation function is very poorly represented by a Gaussian function, which 
only reproduces very low frequencies that do not significantly contribute to the excitation, 
then it fails and underestimates $\chi_k$ by many order of magnitudes (see Sect.~\ref{kumar}). 

The results presented in Fig.~\ref{chik_simu_vs_analytique} are for the depth $r \approx 0.8 R_\odot$, 
where excitation is dominant, and for an angular degree corresponding to the maximum of the kinetic 
energy spectrum ($\ell=40$), whose  contribution is dominant in the excitation rates. 
Those results do not depend on the shell considered but instead on the wavenumber. For very high angular degree 
($\ell > 300$) we find that $\chi_k$ becomes more and more Gaussian. Nevertheless, as shown by Fig.~\ref{spectre}, 
those contributions are negligible compared to large-scale ones.

The value of the parameter $\lambda$ is also of interest. Contrary to the upperlayers where 
$\lambda=1$ \citep{Samadi02I}, we find a higher value, $\lambda=3$, that accords with 
the result of \cite{Samadi02I} who find that the deeper the layers, the higher this parameter.

\subsubsection{The source function}
\label{source}
\begin{figure}[t]
\begin{center}
\includegraphics[height=6cm,width=9cm]{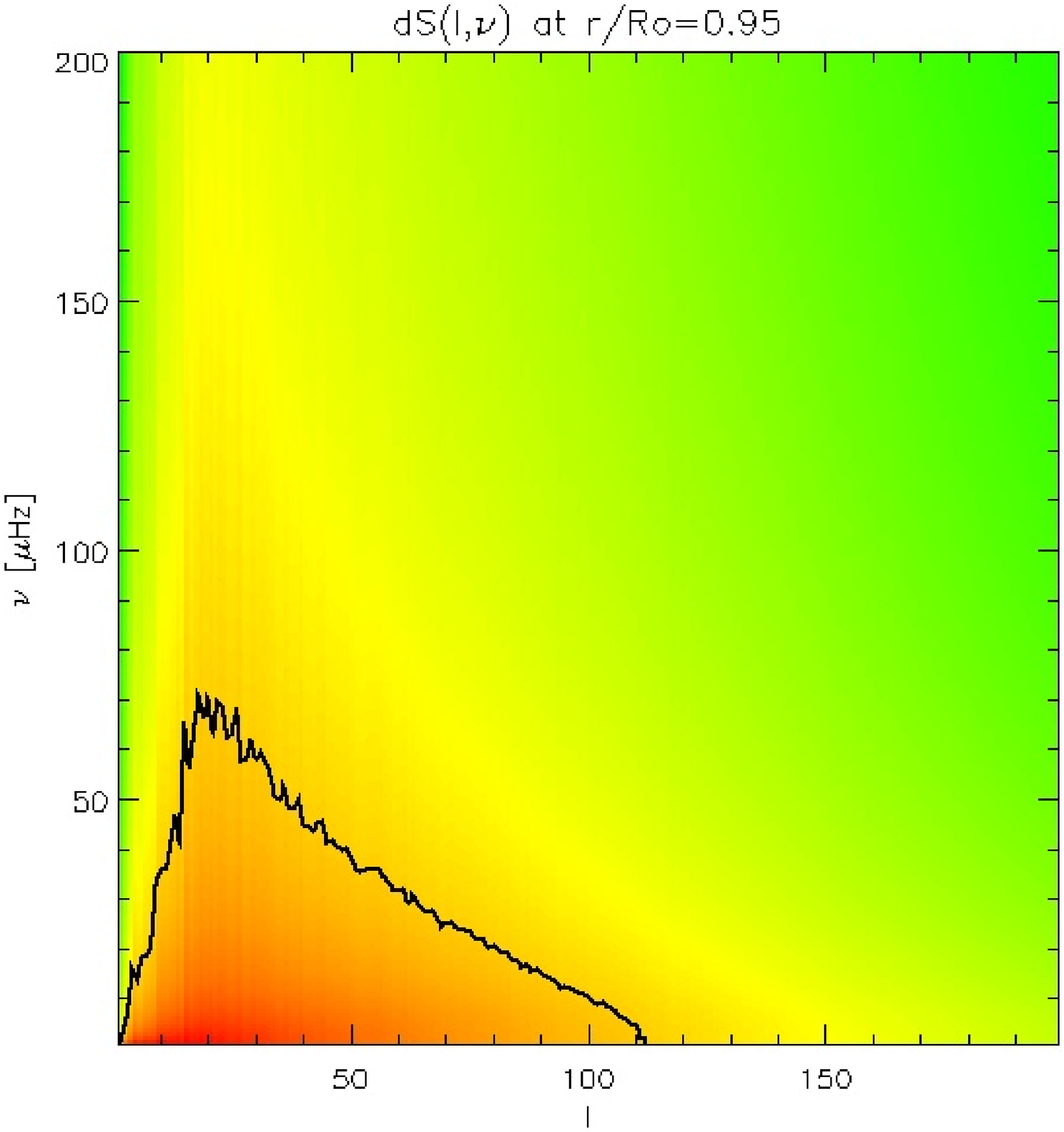}
\includegraphics[height=6cm,width=9cm]{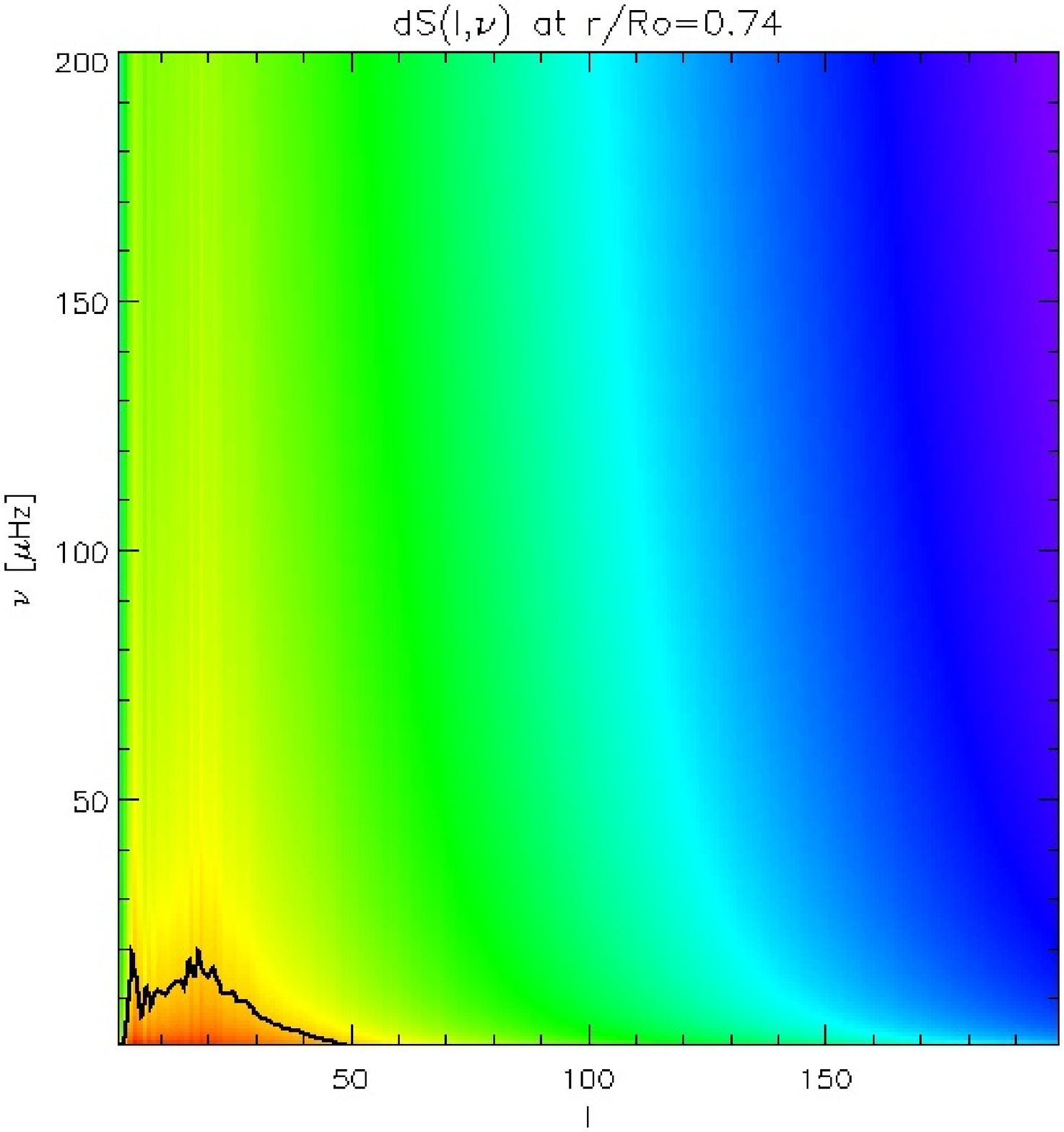}
\caption{The source function is plotted versus the spherical angular degree ({\cal \it l}), and 
the frequency for two radii: $r=0.95 R_\odot$ (top panel) and $r=0.74 R_\odot$ (bottom panel). 
Bright (red) and dark (blue) tones indicate the high and low intensity of the source function, respectively. 
The color table is logarithmic. The black line corresponds to an arbitrary contour line that is  
the same for both panels.}
\label{plot_couleur_fct_source}
\end{center}
\end{figure}

Figure \ref{plot_couleur_fct_source} displays the source function ($\mathcal{S}_k$, \eq{puissance}) 
as a function of both the angular degree {\cal \it l} involved in the summation \eq{Vp} and the mode frequency. The function $\mathcal{S}_k$ evaluated at  two levels, $r=0.95 R_\odot$ and $r=0.74 R_\odot$, is shown in order to emphasize the dependence of $\mathcal{S}_k$ with the radius.  Near the top of the  convection zone, $\mathcal{S}_k$ 
is  non-negligible at high frequencies ($\nu > 50 \mu$Hz) and on small scales. From top to bottom, the intensity 
of  the source function decreases  such that at the bottom, significant intensities exist only  
on large scales (small {\cal \it l} values) and low frequencies. 
 This behavior corresponds to the evolution 
of convective elements, i.e  turbulent eddies evolve on larger time and spatial scales with depth. Thus, 
we conclude that {\it high-frequency $g$ modes are mainly excited in the upper layers, whereas low ones are excited deeper};   however, the net excitation rate,  \eq{puissance}, is a balance between the eigenfunction shape and the source function.

\subsection{Excitation rates}

\begin{figure}[t]
\begin{center}
\includegraphics[height=6cm,width=9cm]{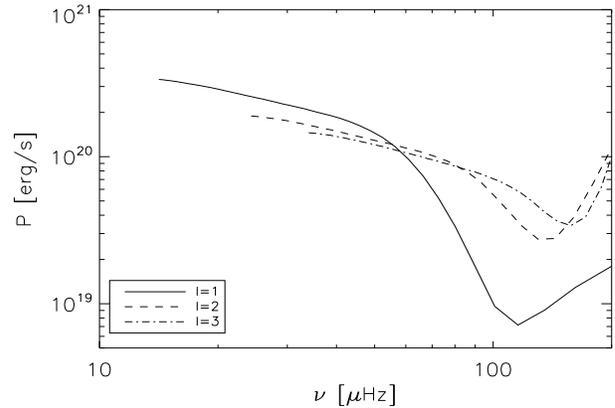}
\caption{Rate ($P$) at which energy is supplied to the modes versus the frequency for modes with
angular degree $\ell=1,2$, and $3$. The computation 
is performed as detailed in Sect.~\ref{compute_power}, using a Lorentzian eddy-time 
correlation function.}
\label{excitation_rates}
\end{center}
\end{figure}

Anticipating the following (see Sect.~\ref{damping}), we stress that modes with high angular degree 
will be highly damped, making their amplitudes very small; hence, we  
restrict our investigation to low-$\ell$ degrees ($\ell < 4$).  
In Fig~\ref{excitation_rates}, we present the excitation rates for low-frequency gravity modes 
(i.e., $\ell=1,2,3$).  By asymptotic modes we denote low-frequency modes  
($\nu < 100\, \mu$Hz, i.e high-$\vert n \vert$ modes) 
while high frequencies ($\nu > 100\, \mu$Hz) correspond to low-$\vert n \vert$ modes.
At  low frequencies ($\nu < 100 \, \mu$Hz), the excitation rate ($P$) 
decreases with increasing $\nu$, it reaches a minimum  and then   at high frequency increases  with the frequency. 
This can be explained by considering the two major contributions to the excitation rate $P$ (\eq{puissance} and \eq{gammabeta})
which are 
the inertia $I$ (in Eq. (1)) and mode compressibility ($\nabla\cdot\xi$, appearing in $R(r)$, Eq. (3)).

\begin{figure}[t]
\begin{center}
\includegraphics[height=6cm,width=9cm]{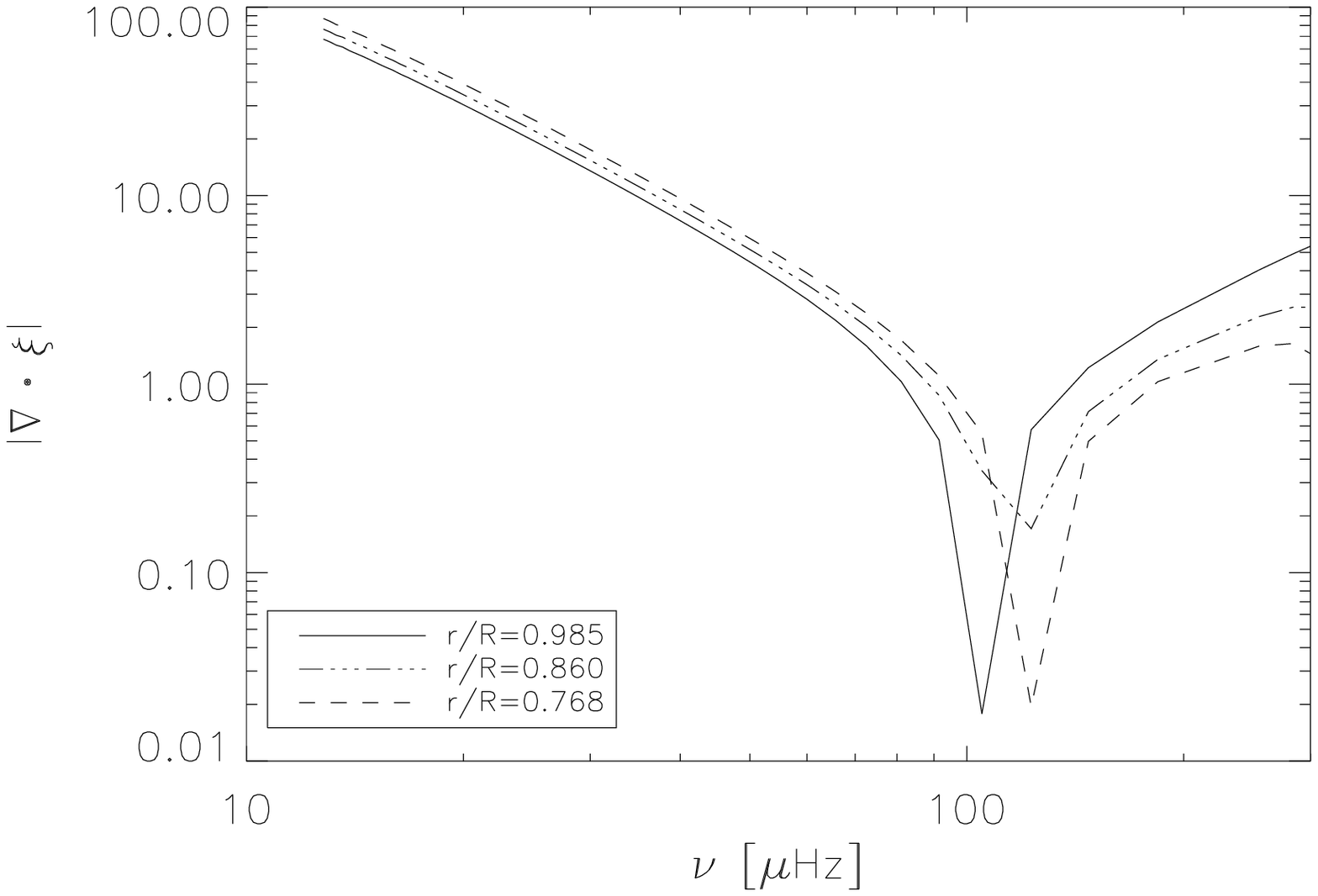}
\includegraphics[height=6cm,width=9cm]{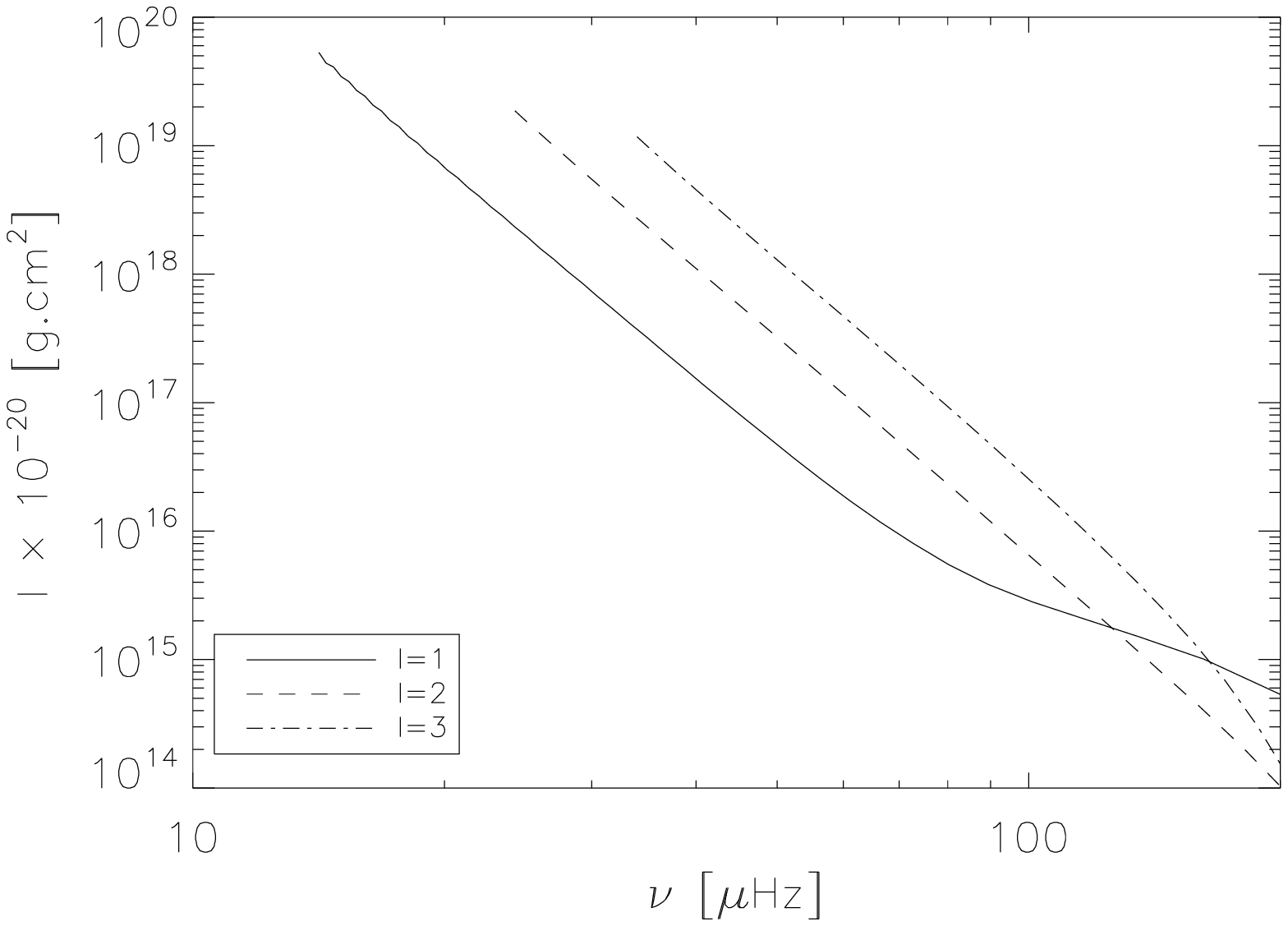}
\caption{{\bf Top:} Absolute value of mode compressibility for $\ell=1$ modes versus the frequency, computed 
for three different layers in the convection zone. 
{\bf Bottom:} Mode inertia versus frequency for modes with  angular degree $\ell=1,2,3$.}
\label{inertie_div}
\end{center}
\end{figure}

Mode inertia decreases with frequency as shown by Fig.~\ref{inertie_div} since 
the higher the frequency, the higher up the mode  is confined in the upper layers. 
This then tends to decrease the efficiency of the excitation of low-frequency modes. 
On the other hand,  mode compressibility (Fig.~\ref{inertie_div}) increases with frequency 
and consequently competes  and dominates the effect of  mode inertia. 
Mode compressibility can be estimated as 
\begin{equation}
\label{div1}
\left \vert \int_\Omega d\Omega ~Y_{\ell}^m ~\vec \nabla \cdot \,  \vec \xi \right \vert \approx \left \vert \deriv{\xi_r}{r} - \ell (\ell + 1) \frac{\xi_h}{r}
\right \vert  
\end{equation}
The mode compressibility 
is minimum when both terms in \eq{div1} are of the same order. 
Following \cite{Belkacem08}, one has
\begin{equation}
\label{dim}
\left \vert \deriv{\xi_r}{r} /  \frac{\ell (\ell+1) \xi_h}{r} \right \vert \simeq \frac{\sigma^{4}}{\ell(\ell+1)} \quad
\textrm{with} \quad  \sigma^2 = \frac{R^3}{GM} \omega_0^2
\end{equation}
where $\sigma$ is the dimensionless frequency, $\omega_0$ is the angular frequency of the mode, 
$R$ the Sun radius, and $M$ its mass. 
According to  \eq{dim},  mode compressibility is minimum   for $\nu \approx 100 \, \mu$Hz depending on $\ell$, as shown by 
Fig.~\ref{inertie_div}. In contrast, in the asymptotic regime ($\nu < 100 \, \mu$Hz), the modes are 
compressible and this compressibility increases with decreasing frequency.

It is important to stress that for the asymptotic $g$~modes, in the frequency range $[20;110]\,\mu$Hz, 
the horizontal contributions in \eq{gammabeta} are dominant. 
For low-$\ell$ $g$ modes, the dominant contributions come, 
in \eq{gammabeta}, from the component of the mode divergence (see \eq{div1}). 
%Such terms correspond 
%to the diagonal part of the tensor $\vec \nabla : \vec \xi$ \citep[see][for details]{Belkacem08}. 
Then 
the ratio of the horizontal to the vertical contributions to \eq{puissance} is around a factor five, imposing the use of a non-radial formalism.


\section{Damping rates}
\label{damping}
To compute theoretical (surface velocities) amplitudes  of $g$~modes, 
 knowledge of the damping rates is required.

\subsection{Physical input}
\label{compute_dampings}

Damping rates have been computed with the non-adiabatic 
pulsation code MAD \citep{MAD02}. This code includes a time-dependent convection 
(TDC) treatment described in \cite{MAD05}: it takes into account the role played 
by the variations of the convective flux, the turbulent pressure, and the dissipation rate of 
turbulent kinetic energy. 
This TDC treatment is non-local, with three free parameters $a$, 
$b$, and $c$ corresponding to the non-locality of the convective flux, the turbulent pressure 
and the entropy gradient. We take here the values $a=10$, $b=3$, and $c=3.5$ obtained by 
fitting the convective flux and turbulent pressure of 3D hydrodynamic simulations in the upper 
overshooting region of the Sun \citep{MAD06a}. According to \cite{MAD05}, we 
introduced a free complex parameter $\beta$ in the perturbation of the energy closure equation. 
This parameter is introduced to prevent non-physical spatial oscillation of the eigenfunctions. 
%because of the lack of knowledge. 
We use here the value $\beta=-0.5i$, which leads 
to a good agreement between the theoretical and observed damping rates and phase lags in the 
range of solar pressure modes \citep{MAD06c}. 
The sensitivity of the damping rates to $\beta$ is discussed in Sect.~\ref{discussion_beta}, and we 
show in next sections that the values of those parameters have no influence on the results 
since we are interested in low-frequency $g$ modes.

We use the TDC treatment as described in \cite{MAD06b}, 
in which the 1D model reproduces exactly 
the mean convective flux, the turbulent pressure and the mean 
superadiabatic gradient obtained from a 3D hydrodynamic simulation by \cite{Stein98}, by 
introducing two fitting parameters, the mixing-length, and a closure parameter 
\citep[see][for details]{MAD06b}.
We also stress that, for low-frequency $g$~modes, particular attention is to be paid to the 
solution of the energy equation near the center as explained in Appendix~\ref{enregy_eq} 
for the $\ell=1$ modes since those dipolar modes present a peculiar behavior near the center 
that must be properly treated.  

\subsection{Numerical results for a solar model}

\subsubsection{Sensibility to the time-dependent treatment of convection}
\label{discussion_beta}

\begin{figure}
\begin{tabular}{c}
\includegraphics[height=6.5cm,width=9cm]{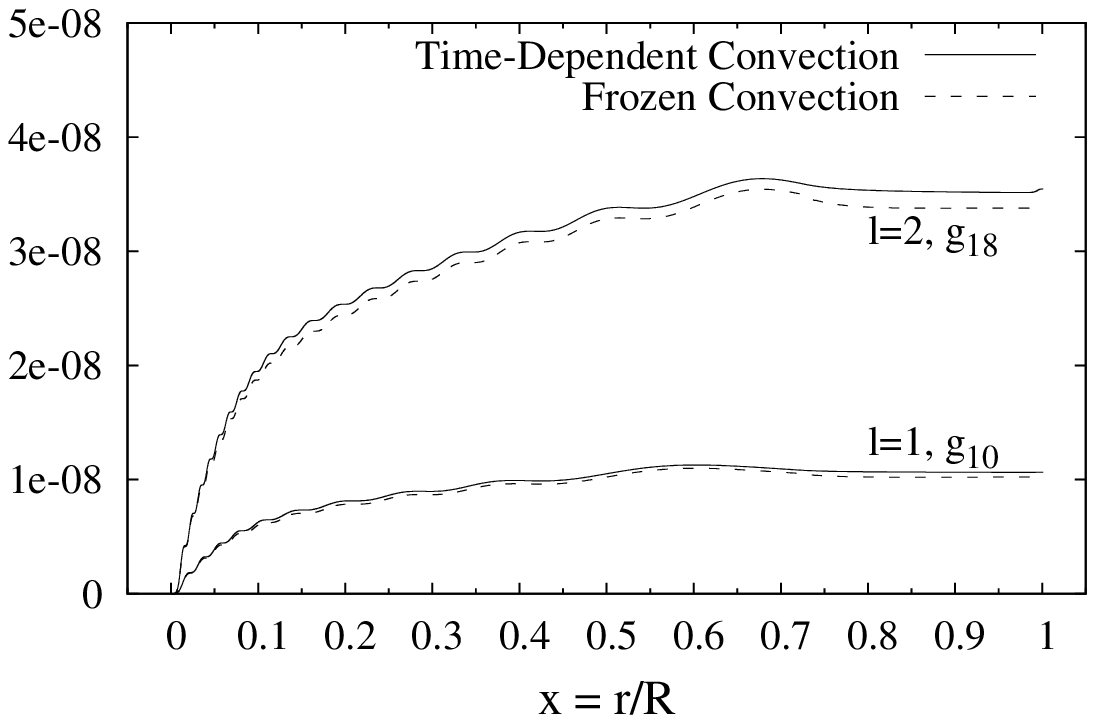} \\[-0.3cm]
\includegraphics[height=6.5cm,width=9cm]{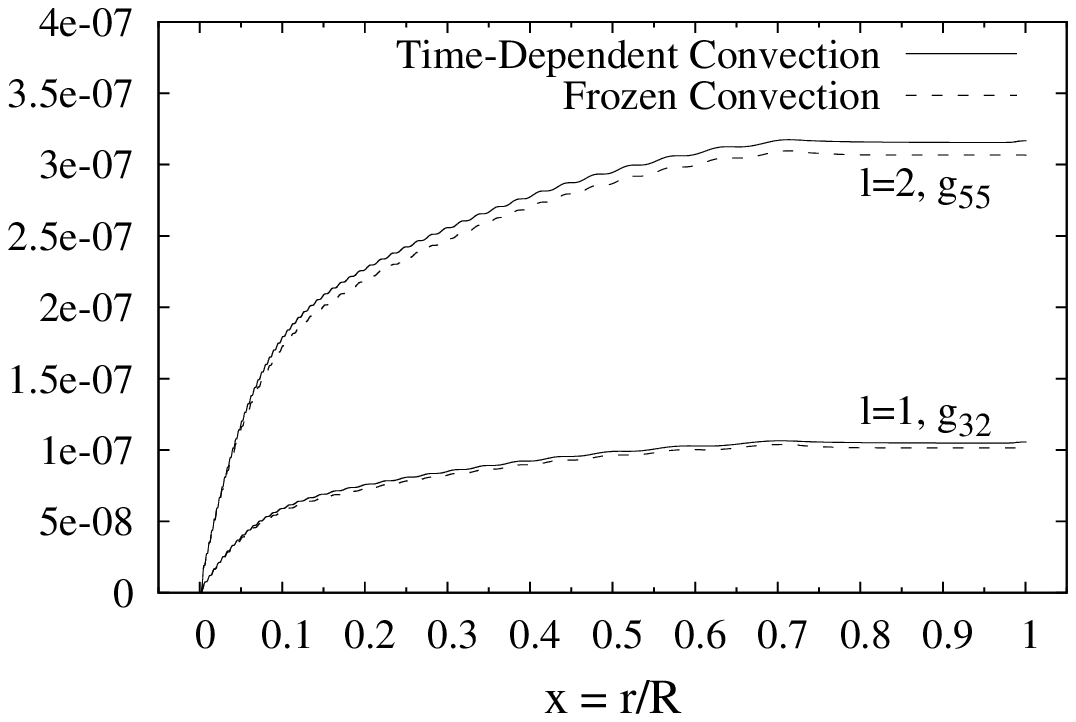}
\end{tabular}
\caption{Work integrals  for $\ell=1$ and $\ell=2$ modes at $\nu\simeq60\mu$Hz 
(top panel) and  $\nu\simeq20\mu$Hz (bottom panel), the surface values give the damping 
rates $\eta$ in $\mu$Hz.}
\label{workfig}  
\end{figure} 

To understand the contribution of each layer of the Sun in the damping of the $g$ modes, 
we give the normalized work integral in Fig.~\ref{workfig} in such a way that the 
surface value is the damping rate $\eta$ (in $\mu$Hz)\footnote{
Note that, regions where the work decreases outwards have a damping effect 
on the mode, or a driving effect when it increases outwards.}.
Results obtained with our TDC treatment (solid lines) and with 
frozen convection (FC, dashed line) are compared for 4 different modes with 
$\nu\simeq60\mu$Hz (top panel) and  $\nu\simeq20\mu$Hz (bottom panel). 
We see that most of the damping occurs in the inner part of the radiative core. 
The work integrals obtained with TDC and FC treatments are not very different;  
hence, the uncertainties inherent in the treatment of the coherent interaction 
between convection and oscillations do not significantly affect the theoretical 
damping rates of solar asymptotic $g$ modes. 
This means that the frozen convection is adapted to low-frequency $g$~modes.  
This can be explained by paying attention to the ratio 
$\mathcal{Q}=\omega_0 / \omega_c$, 
where $\omega_0$ is the oscillation frequency and $\omega_c$ the convective 
frequency, defined to be $\omega_c = 2\pi \Lambda / u_{mlt}$ where $\Lambda$ is the mixing length and 
$u_{mlt}$ the convective velocity. In the whole solar convective zone $\mathcal{Q}$ is higher than unity 
except near the surface (the superadiabatic region). 
However  contributions   of  the surface layer
remain small in comparison with the radiative ones for 
asymptotic $g$ modes (see Fig.~\ref{workfig}).

One can thus draw some conclusions
\begin{itemize}
\item for high-frequency $g$ modes ($\nu > 110 \, \mu\textrm{Hz}$), 
the work integrals and thus the damping rates are sensitive to the parameter $\beta$ 
that is introduced to model the convection/pulsation interactions
%the latest conclusion is no longer valid 
because the role of the surface layers in the work integrals becomes important. As a result, 
the results on the damping rates are questionable for high frequencies since the 
value of $\beta$ is derived from the observed $p$ modes and that there is no 
evidence it can be applied \emph{safely}  for $g$ modes.
\item in contrast,  for low-frequency $g$~modes ($\nu < 110 \, \mu\textrm{Hz}$), we find that 
the work integrals and then the damping rates are {\it insensitive} to parameter $\beta$.  
%that is introduced to model the convection/pulsation interactions. 
Also, we numerically checked that the damping rates are insensitive to the 
non-local parameters introduced in Sect.~\ref{compute_dampings}. 
\end{itemize}

\begin{figure}
\begin{tabular}{c}
\includegraphics[height=6.5cm,width=9cm]{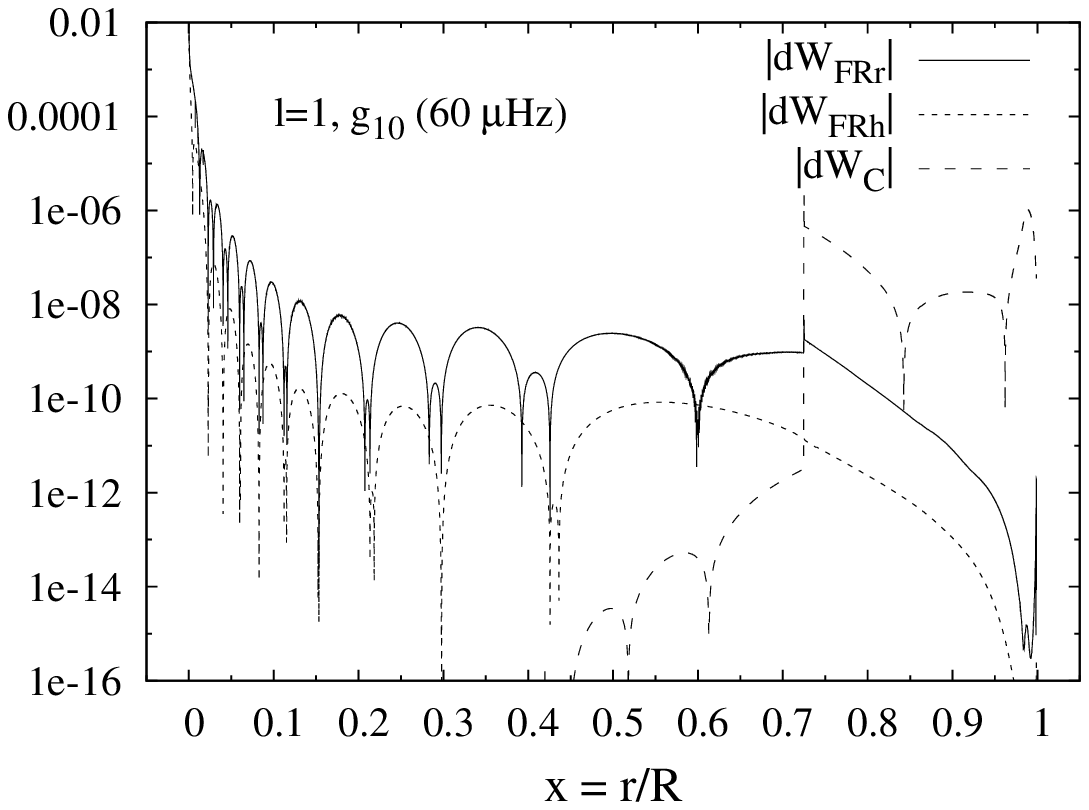} \\[-0.3cm]
\includegraphics[height=6.5cm,width=9cm]{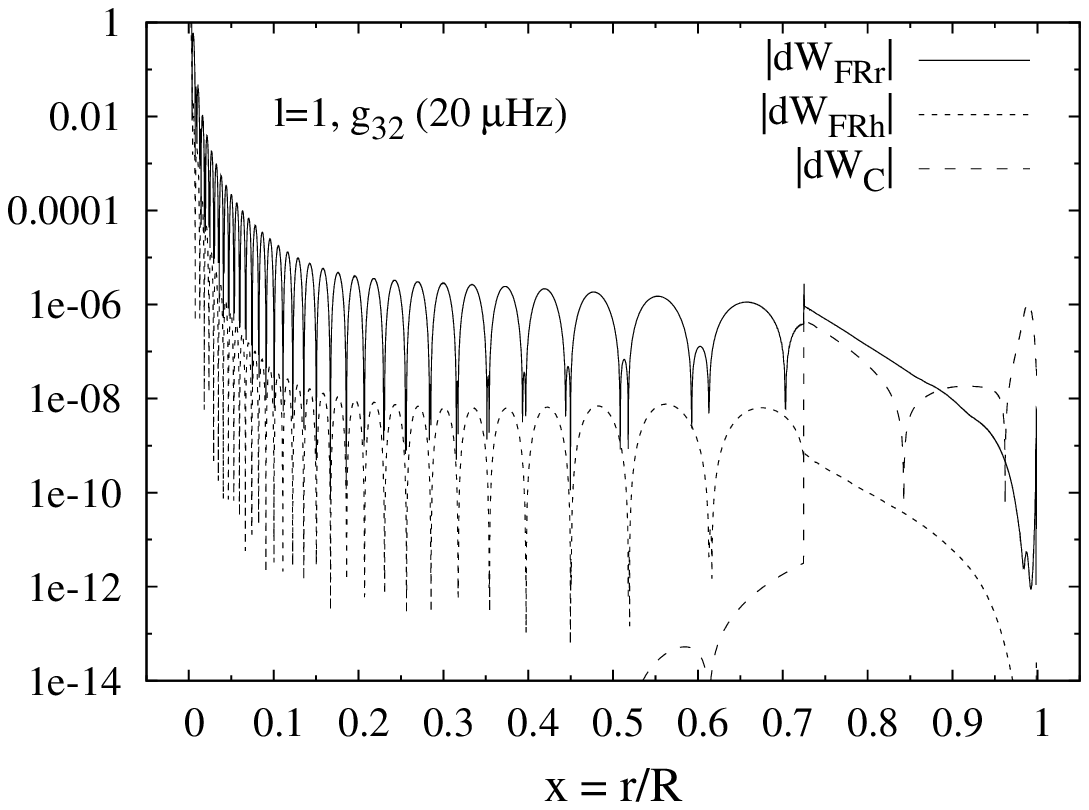}
\end{tabular}
\caption{Contributions to the work by the radial radiative flux variation (solid line), 
the transverse radiative flux variation (dotted line), and the time-dependent convection 
terms (dashed line), for the mode $\ell=1$, g$_{10}$ (top panel) and $\ell=1$, g$_{32}$ 
(bottom panel). Details are given in the text.}
\label{contribfig}  
\end{figure} 

\subsubsection{Contributions to the work integral}

Figure~\ref{contribfig} allows us to investigate the respective roles played by different terms in the damping 
of the mode. More precisely we consider two modes  ($\ell=1$, g$_{10}$, 
$\nu \approx 60 \, \mu$Hz and $\ell=1$, g$_{32}$ , $\nu \approx 20 \, \mu$Hz)
 in the frequency interval of interest here and  give in Fig.~\ref{contribfig}  the modulus of:

\begin{itemize}
\item the contribution to the work by the radial part of the radiative flux 
divergence variations (solid line)
\begin{equation}
\label{contrib_rad}
dW_{FRr}=\Re\{ \left({\delta T \over T}\right)^*\,{\partial\delta L_R\over \partial x} \}\:  {R\over G M^2\sigma} \; , 
\end{equation}
where $T$ is the temperature, $L_R$  the radiative luminosity, $R$ the solar radius, 
$M$ the solar mass, $x$ the normalized radius, $\sigma$ the real part of the normalized frequency  
$\sigma = \omega_0/(GM/R^3)^{1/2}$, 
and $x$ the normalized radius (see Appendix~\ref{enregy_eq}). 
Note that $\delta$ denotes the wave Lagrangian perturbations, $\Re$ the real part, 
and $^*$ the complex conjugate. 
\item the contribution to the work by the transversal part of the radiative 
flux divergence variations (dotted line): 
\\
\begin{equation}dW_{FRh}=-\ell(\ell+1)\:\Re\left\{\frac{\delta T^*}{T}\left(\frac{\delta T}{x dT/dx} 
- \frac{\xi_r}{r}\right)\right\}\:\frac{R L}{G M^2\sigma x} \, ;
\end{equation}
\item the contribution to the work by the time-dependent convection 
terms (dashed line): $dW_{C}$ \citep[see Sect. 4 of][]{MAD05}. 
%Despite the dominance of the $g_{10}$ mode in the convective region in Fig.~\ref{contribfig}, 
%Fig.~\ref{workfig} shows that this contribution is globally negligible.  
\end{itemize}
Integration of these terms over the normalized radius gives their global contribution 
to the work performed during one pulsation cycle. 

The time-dependent convection terms have a very low weight for both 
modes in the frequency range $\nu < 110 \, \mu\textrm{Hz}$. 
It confirms the conclusion of Sect.~\ref{discussion_beta} that the damping rates of low-frequency 
$g$~modes 
are not dominated by the perturbation of the convective flux,\emph{i.e.} 
the interaction convection/oscillation (through the parameter $\beta$). 
The higher the mode frequency, the higher the integrated convective contribution of the work ($W_{C}$), 
which becomes dominant for $\nu > 110 \, \mu\textrm{Hz}$.

While the transverse radiative flux term plays a significant 
role near the center, the major contribution to the work comes from the 
radial component of the radiative flux variations. As a result, the radiative damping 
is the dominant contribution for low-frequency gravity modes. 

\begin{figure}
 \resizebox{\hsize}{!}{\includegraphics{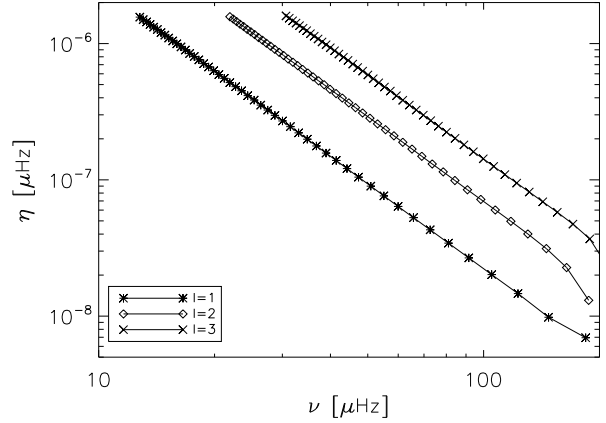}}
 \caption{Theoretical damping rates $\eta$ of g modes of degree $\ell=1, 2, 3$ 
 as a function of the oscillation frequency in $\mu$Hz.}
 \label{etafig}
\end{figure} 

In Fig.~\ref{etafig}, we give the theoretical damping rates $\eta$ of g-modes of degree 
$\ell=1, 2, 3$, as a function of the oscillation frequency in $\mu$Hz. We see that for 
$\nu < 110 \, \mu$Hz, $\eta$ is a decreasing function of frequency. 
We find that the frequency dependence is $\eta \propto \omega_0^{-3}$. To understand 
this behavior, we express the integral expression of the damping rate 
\citep[see][for details]{MAD05} as 
\begin{equation}
\label{dampings_radiatif}
\eta = \frac{1}{2 \, \omega_0 I} \int_{0}^{M} 
\mathcal{I}m \left(\frac{\delta \rho}{\rho}^* T \delta S\right) \left(\Gamma_3 - 1\right) \textrm{d}m
\end{equation}
with
\begin{equation}
I = \int_{0}^{M} \textrm{d}m \; | \vec \xi |^2 \quad \textrm{and} \quad  \left(\Gamma_3 - 1\right) 
= \left(\frac{\partial ln T}{\partial ln \rho}\right)_s 
\end{equation}
where $\delta \rho, \delta S$ are the perturbations of the density and entropy, respectively, 
$\rho, T$ are the density and temperature, $\vec \xi$ the eigenfunction, and 
the star denotes the complex conjugate.

Keeping only the radial contribution of the radiative flux in the energy equation (\eq{E} ) 
because it is the dominant contribution, 
and neglecting the production of nuclear energy ($\epsilon=0$), 
one gets
\begin{equation}
\label{deltaS}
T \delta S = \frac{i}{\omega_0} \frac{\partial \delta L}{\partial m} \, .
\end{equation}
This approximation comes from the dominance of the radial contribution of the radiative flux. 
In addition, in the diffusion approximation
\begin{equation}
\label{deltaL}
\frac{\delta L}{L} =  \left( 
\frac{1}{\left(\textrm{d}T / \textrm{d}r\right)}\frac{\partial \delta T}{\partial r} 
+2\,  \frac{\xi_r}{r} + 3\, \frac{\delta T}{T} 
- \frac{\delta \kappa}{\kappa} - \frac{\delta \rho}{\rho} 
- \frac{\partial \xi_r}{\partial r}\right)
\end{equation}
Because of the high wavenumber for low-frequency $g$ modes, the term in $\partial\delta T/\partial r$ is very high 
in  \eq{contrib_rad}, dominates in Eq.~(\ref{deltaL}), and is the main source of damping. This term appears 
as a second-order derivative in the work integral, and introduces a factor $k_r^2$ 
($k_r \approx \sqrt{\ell(\ell+1)}N/(\omega_0 r)$ is the vertical local wavenumber). 
Thus, from Eqs. (\ref{deltaL}), (\ref{deltaS}), and (\ref{dampings_radiatif}) one obtains  
$\eta \propto \omega_0^{-5} / I$. 
By using an asymptotic expansion of the eigenfunctions \citep{JCD02}, one gets $I \propto \omega_0^{-2}$, 
which permits $\eta \propto \omega_0^{-3}$ and explains 
the behavior of $\eta$ in Fig.~\ref{etafig}. 
The argument is the same for the variation of $\eta$ with the angular degree at fixed frequency 
becomes it comes from the wave-number dependence $k_r^2$.

Above ~$110 \, \mu$Hz, the role of the radiative zone in the mode damping is smaller. 
There, the damping rates begin to increase with frequency simply because the kinetic energy 
of the modes decreases faster than the mechanical work.

\section{Surface velocities of $g$~modes}

\subsection{Theoretical (intrinsic) velocities}
\label{intrinsic}

We compute the mean-squared surface velocity ($v_s^2$) for each mode as
%according to the relation \citep{Baudin05}:
\begin{align}
\label{Pobs0}
v_s^2 (h) &=\left< \int_\Omega  \Bigl(\vec v(\vec r,t) \cdot \vec v(\vec r,t)\Bigr) d\Omega \right> (h)
\end{align}
where $h$ is the height in the stellar atmosphere, $<>$ the time average. 
Using the expression \eq{vitesse_annexe} in appendix C, one then has
\begin{align}
\label{Pobs}
%v_s^2 (h) =  v_r(h)^2 + \ell (\ell+1) v_h(h)^2\\
v_s^2 (h) &= A^2 \left[ v^2_r(h) + \ell (\ell+1)  v^2_h(h) \right] \, .
\end{align}
The amplitude $A^2=(1/2)\left<|a(t)|^2\right>$ is given by (\eq{amp}):
\begin{align}
A^2 = \frac{ P}{2 \, \eta \, I  \omega_0^2}
\end{align}
where $<>$ denotes the time average, 
$I$ the mode inertia, $ \eta $ the damping rate, and
$v_{r,h}(h) = \omega_0 \, \xi_{r,h}(h)$ with $\xi_r(h)$ and $\xi_h(h)$  
respectively the radial and horizontal displacement eigenmode components. 

%Note however, as shown by \cite{Berthomieu90}, the results on the surface 
%velocities are only slightly sensitive to the height ($h$).

In this section,
 we consider the level of the photosphere $h=R$ with $R$ the radius
  at the photosphere. Figure~\ref{fig_comp_kumar} presents \emph{intrinsic} 
 values of the velocities. 
%For asymptotic $g$~modes 
%presented in Fig.~\ref{vitesses}, the horizontal component of the eigenfunction is dominant.
The behavior of the surface velocities as a function of the angular degree ($\ell$) is mainly due to 
the damping rates, which rapidly increase with $\ell$; hence, at fixed frequency,  the 
higher the angular degree, the lower the surface velocities. As a consequence, amplitudes 
are very low for $\ell >3$ . 
At fixed $\ell$, $v_s$ increases with frequency with a slope resulting from 
a balance between the excitation  and  damping rates. 
Nevertheless, modes of angular degree $\ell=1$ exhibit a singular behavior, i.e. a maximum at 
$\nu \approx 60 \, \mu$Hz. This is due to the variation of the slope in the excitation rates 
(see Fig.~\ref{excitation_rates}). 
In terms of amplitudes, the maximum is found to be $\approx 5 \,$mm/s for $\ell=1$ at 
$\nu \approx 60 \,\mu$Hz, which corresponds  to the mode with radial order $\vert n \vert = 10$. 
It is important to stress that the velocities shown  in Fig.~\ref{fig_comp_kumar}, are \emph{intrinsic} 
values of the modulus that must not be confused with the \emph{apparent} 
surface velocities (see Sect.~\ref{apparent}), which are the values that can be compared with  observed ones. 

\subsection{Comparison with previous estimations}
\label{discussion}

The theoretical intrinsic velocities  obtained in the present  work must be compared to previous estimations 
based on the same  assumption that modes are stochastically excited by turbulent convection. 
All works cited in the next sections deal with intrinsic velocities, \emph{i.e.} ones not corrected for  visibility effects.

\subsubsection{Estimation based on the equipartition of energy}

The first estimation of $g$-mode amplitudes was performed by \cite{Gough85},  
who found a maximum of velocity of about $0.5 \, \textrm{mm\,s}^{-1}$ for 
the $\ell=1$ mode  at $\nu \approx 100\, \mu$Hz. 
\cite{Gough85} used the principle of equipartition of energy, which  consists in equating the mode energy  
(${\cal E}$) with the kinetic energy 
of resonant eddies whose lifetimes are close to the modal period. This "principle" has been 
theoretically justified for $p$ modes, by \cite{GK77} assuming that the modes are damped by eddy viscosity. 
They found that the modal energy to be inversely proportional to the damping rate, $\eta$, and 
proportional to an integral 
involving the term  
$E_\lambda \, v_\lambda \, \lambda$ where $E_\lambda \equiv
(1/2)\, m_\lambda \, v_\lambda^2$ is the kinetic
  energy of an eddy with size $\lambda$, velocity $v_\lambda$, and mass 
  $m_\lambda=\rho \, \lambda^3$ \citep[see Eq.~(46) of][]{GK77}. 
Using a solar model, they show  that the damping rates  of solar $p$ modes are dominated by turbulent viscosity
and that  the damping rates are  accordingly  proportional to the
eddy-viscosity, that is, $\eta \propto v_\lambda \, \lambda$ \citep[see Eq.~(6) of][]{GK77a}. 
Hence, after some simplifying manipulations,   \cite{GK77} found the
modal energy to be (see their Eq.~(52)) 
\begin{align}
\mathcal{E} \approx 0.26 \, E_\lambda  = 0.13 \, m_\lambda \,
v_\lambda^2 \;.
\end{align}

This principle then was  used by \cite{CD83} for $p$ modes and \cite{Gough85} for solar $g$ modes. 
However, as mentioned above,  the result strongly depends on the way the modes are damped, and  
for asymptotic $g$ modes there is no evidence that this approach can be used and in particular, 
as shown in this work, if the damping is dominated by radiative losses. 

\subsubsection{\cite{Kumar96}'s formalism}
\label{kumar}

Another study was performed by \cite{Kumar96}, which was motivated by a claim 
of $g$-mode detection in the solar wind \citep{Thomson95}. 
Computations were performed using the \cite{GK94} formalism;  
both turbulent and radiative contributions to the damping rates 
were included as  derived by \citep{Goldreich91} who 
obtained mode lifetimes around $10^6 \, \textrm{yrs}$.  
This is not so far from our results (see Fig~\ref{etafig}). 
 \cite{Kumar96} found that the theoretical (i.e. not corrected for visibility factors) surface velocity is around 
$10^{-2} \, \textrm{cm}\,\textrm{s}^{-1}$ near $\nu = 200 \, \mu\textrm{Hz}$ for  $\ell=1$ modes. 
However, as shown in Sect.~\ref{damping},  the results for this frequency range are very sensitive 
to the convective flux perturbation in the damping rate calculations. 
Thus, we do not discuss the result obtained for those frequencies. 

More interesting for our study,  \cite{Kumar96} also found very low velocities 
($10^{-2}\,  \textrm{mm}\,\textrm{s}^{-1}$) for $\nu < 100 \, \mu\textrm{Hz}$. This is significantly lower than what we find. 
However, the efficiency of the excitation strongly depends on how the eddies and 
the waves are temporally-correlated. As already explained in Sect.\ref{compute_power}, the way the eddy-time correlation function is modeled 
is crucial since it leads to major differences between, for instance, a Gaussian and a Lorentzian 
modeling. 
The \cite{GK77} approach, from which \cite{Kumar96}'s formulation is derived, 
implicitly assumes that the time-correlation between the eddies is Gaussian.   
The present work  (as explained in Sect.~\ref{compute_power}) 
assumes a  Lorentzian for the time correlation function  $\chi_k$, which results in $v=3$ mm/s 
in amplitude for $\ell=1$ mode at $\nu \approx 60\,\mu$Hz (Sect.~\ref{intrinsic}). 

\begin{figure}[t]
\begin{center}
\includegraphics[height=6cm,width=9cm]{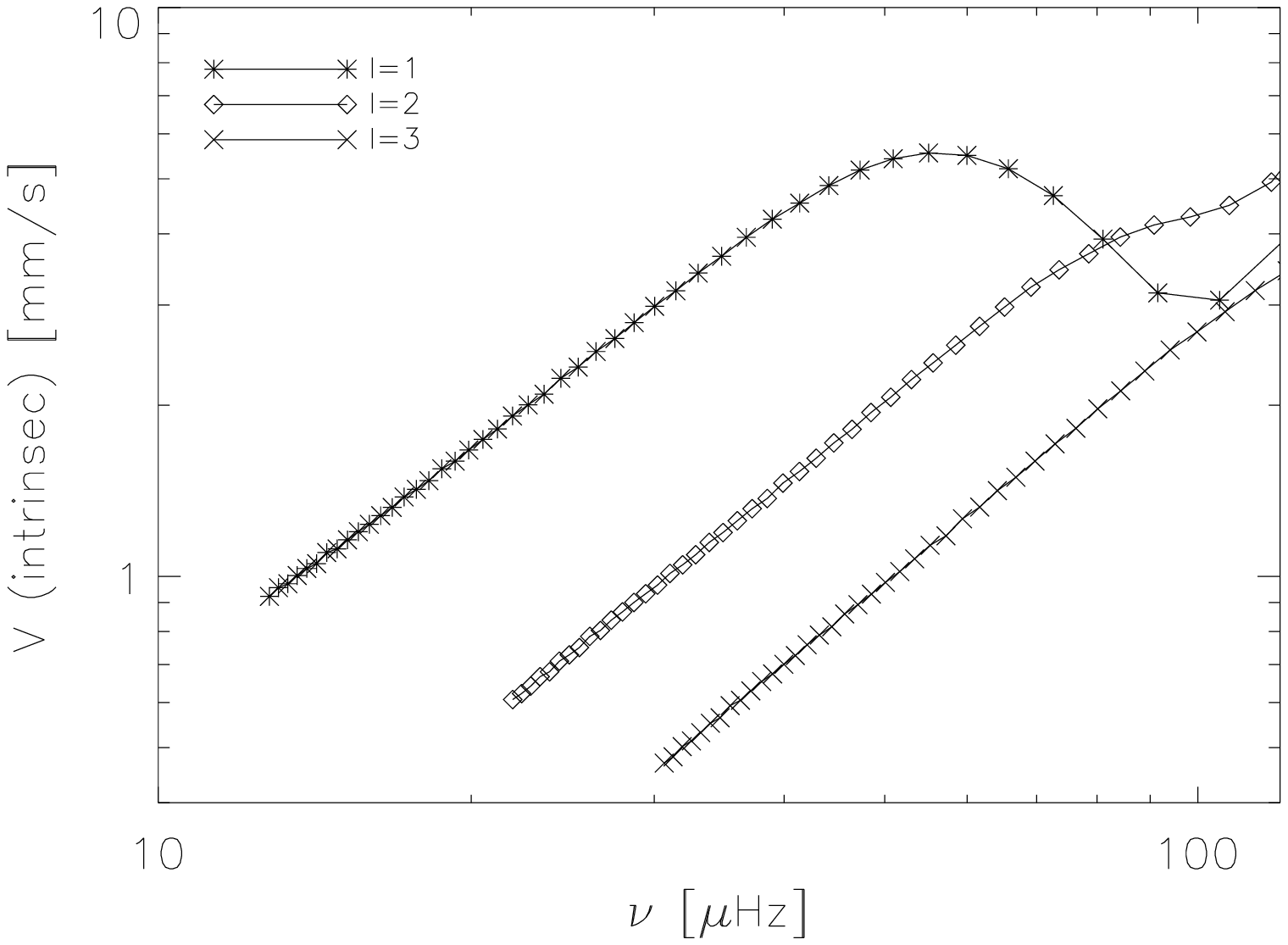}\\
\includegraphics[height=6cm,width=9cm]{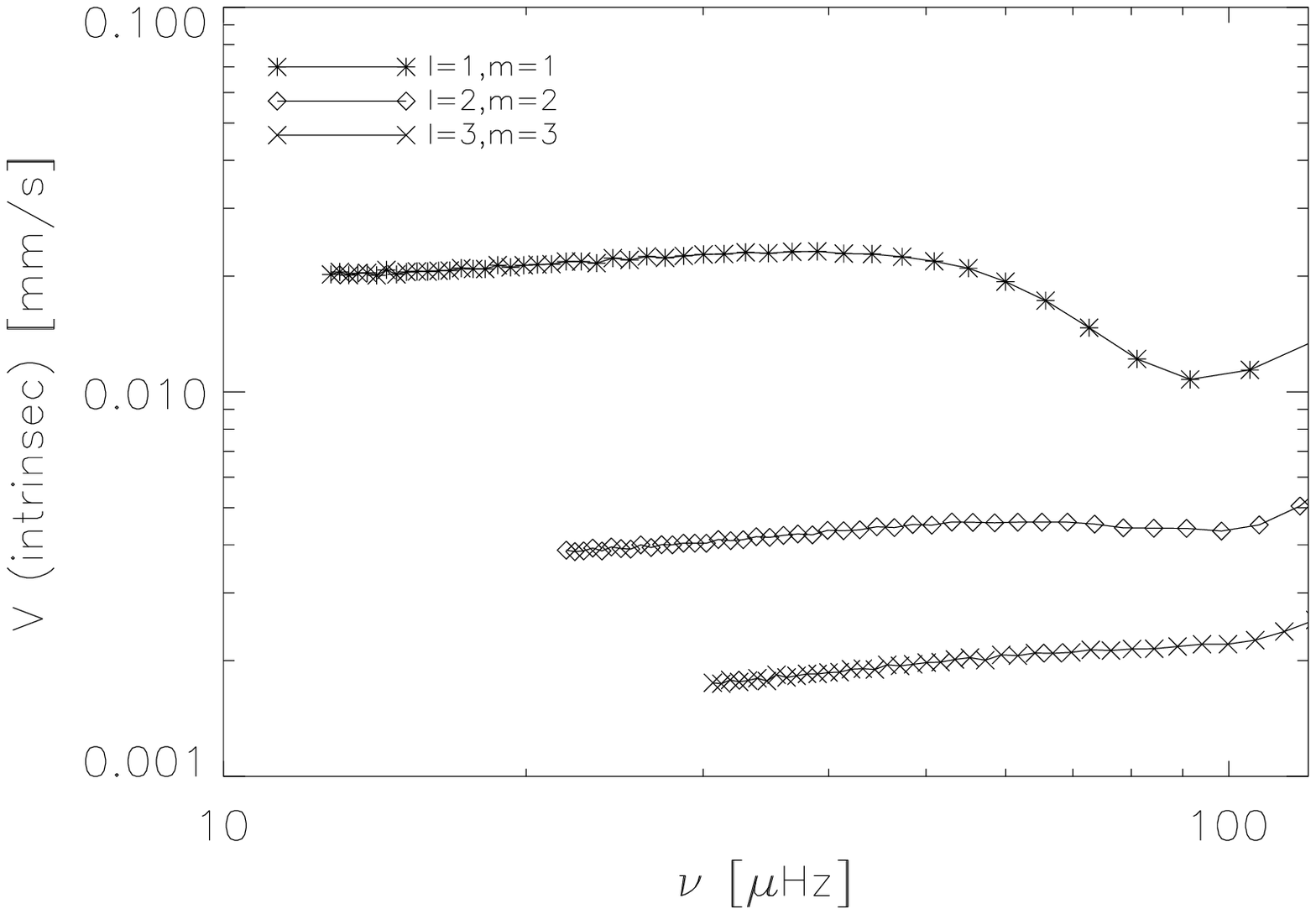}
\caption{{\bf Top:} Theoretical \emph{intrinsic} surface velocities of g-modes of degree $\ell=1, 2, 3$ 
as a function of the oscillation frequency in $\mu$Hz, computed as described  
in Sect.~\ref{intrinsic} using a \emph{Lorentzian} $\chi_k$.  
{\bf Bottom:} Surface velocities of gravity modes of angular degree $\ell=1$ and $\ell=2$ computed 
using a \emph{Gaussian} $\chi_k$ and a Kolmogorov spectrum to reproduce the results of \cite{Kumar96}.}
\label{fig_comp_kumar}
\end{center}
\end{figure}

We performed the same computation but now assuming  
$\chi_k$ to be Gaussian (\eq{eqn:GF}) and using a Kolmogorov spectrum as in \cite{Kumar96}. 
In that case (see Fig.~\ref{fig_comp_kumar}), we find velocities of the order of $10^{-2}\,mm\,s^{-1}$  
for $\ell=1$ which agree with  the result of \cite{Kumar96}, which is significantly lower 
than when assuming a Lorentzian.

\section{Apparent surface velocities}
\label{apparent}

\begin{table}
\begin{center}
\begin{tabular}{lcccc}
$m$ &   0 &   1 &  2 & 3 \\
\hline \\
$\ell$=1 & 0.117   &    0.675  &  & \\
$\ell$=2 & 0.346   &   0.107   &  0.437 & \\
$\ell$=3 & 0.06 &    0.164   &  0.0552 &    0.184
\end{tabular}
\caption{\label{table1} Values of the visibility coefficient $\alpha_{\ell}^{m}$ of 
the radial component of the velocity, 
corresponding to  an inclination angle of  $\theta_0=83^o$.}
\end{center}
\end{table}
\begin{table}
\begin{center}
\begin{tabular}{lcccc}
$m$ &   0 &   1 &  2 & 3 \\
\hline \\
$\ell$=1 & 0.094   &    0.540  &  & \\
$\ell$=2 & 0.833   &   0.258   &  1.053 & \\
$\ell$=3 & 0.291 &    0.649   &  0.268 &    0.892
\end{tabular}
\caption{\label{table2} Values of the visibility coefficient $\beta_{\ell}^{m}$ of 
the radial component of the velocity, 
corresponding to  an inclination angle of  $\theta_0=83^o$.}
\end{center}
\end{table}

We denote as \emph{disk-integrated apparent velocities} the values of amplitudes that take both geometrical 
and limb darkening effects into account. Contrary to solar $p$~modes, one cannot neglect the 
horizontal component of $\vec \xi$ compared to the vertical one. 
The observed velocity ($V_{obs}$) is given by the apparent surface velocity 
$<|V_{app}(r,t)|^2>^{1/2}$(see Appendix.~\ref{visibilite})  evaluated at the observed  line formation height 
$h$:
\begin{align}
V_{obs} = \left({P\over 2 \eta I \omega_0^2}\right)^{1/2}~\left( \alpha_\ell^m \, v_r(h) + \beta_\ell^m \, v_h(h) \right)
\end{align}
where  $\alpha_\ell^m$ and $\beta_\ell^m$ are the visibility factors defined in Appendix.~\ref{visibilite}.

In Appendix ~\ref{visibilite} we follow the procedure first derived by \cite{Dziembowski77} and 
for asymptotic $g$ modes by \cite{Berthomieu90}. 
We use a quadratic limb-darkening law following \cite{Ulrich00} for the Sun with an angle 
between the rotation axis and the Equator of $83^Á$ degrees. 
 As mentioned above, the apparent velocities are evaluated at the level  $h$, \emph{i.e.} 
the height above the photosphere where oscillations are measured. Then $h$ is set so as to correspond to the  
SoHO/GOLF measurements that use the NaD1 and D2 spectral lines,  
formed at the optical depth $\tau=5.10^{-4}$ \citep[see][]{Bruls92}. 
The results are presented in Tables~\ref{table1} and \ref{table2} for angular degrees 
$\ell=1,2,3$. 

\begin{figure}
\includegraphics[height=6cm,width=9cm]{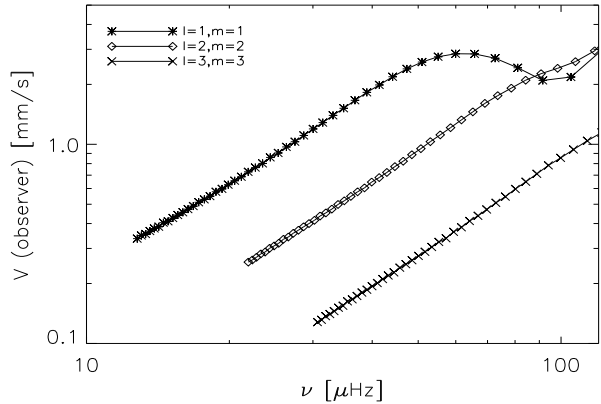}
\caption{
Apparent surface velocities for $g$ modes of degree $\ell=1, 2, 3$  as a function of the oscillation frequency in $\mu$Hz
(visibility factors are taken into account). }
\label{vitesses}
\end{figure}

Figure~\ref{vitesses} displays the apparent velocities for modes $\ell=1,2,3$ and $\ell=m$.  
For a given angular degree, the azimuthal order degree is chosen such that the apparent velocity is maximal. 
The velocities of the $m=0$ modes are strongly attenuated by the visibility effects, while 
the $m=\ell$ modes are less sensitive to them. 
For $\ell=1$ modes, the amplitudes are divided by a factor of two with respect to 
the intrinsic velocities, while the $\ell=2, 3$ mode velocities remain roughly the same.     
Consequently, our calculations show that both the $\ell=1$ and $\ell=2$ ($m=\ell$) are 
the most probable candidates for  detection with amplitudes $\approx 3$ mm s$^{-1}$. 

\subsection{Detectability of $g$ modes; only a matter of time}

To compare our calculated apparent velocities 
with
observations, we used data from the GOLF spectrometer \citep{Gabriel02} 
onboard the SOHO platform, which performed Doppler-like measurements on
the disk-integrated velocity of the Sun, using the Na D lines. We used here
a series of 3080 days to estimate the background noise level and compare it to
the apparent velocities determined in this work.

A first possible approach is to use some analytical and statistical calculations such 
as the ones developed by \cite{Appourchaux00} (Eq.~10). Once a length of
observation $T$ (in units of 10$^6$s), a frequency range $\Delta \nu$ (in
$\mu$Hz), and a
level of confidence $p_{det}$ are set, this gives the corresponding
signal-to-noise ratio
\begin{equation}
\frac{s_{det}}{<s>} \simeq \ln(T) + \ln(\Delta \nu) - \ln(1-p_{det}) \, ,
\label{trucmuche}
\end{equation}
where $s_{det}$ is the power of the signal to be detected, and $<s>$ the local
power of the noise. This means that any peak in the frequency range $\Delta \nu$
above this ratio has a probability $p_{det}$ of not being due to noise.
Choosing a frequency range of $\Delta \nu = 10\mu$Hz centered on the frequency
of the highest expected velocities (60$\mu$Hz) sets the background level at
$\approx$ 500 (m\,s$^{-1})^2/$Hz. 
Equation (\ref{trucmuche}) gives an amplitude
of $5.2~mm\,s^{-1}$ for a detection with a confidence level $p_{det}$ of 90\% for 15
years of observation, $4.6~mm\,s^{-1}$ for 20 years, and $3.8~mm\,s^{-1}$ for 30 years.

\begin{figure}
\includegraphics[height=9cm,width=6cm,angle=90]{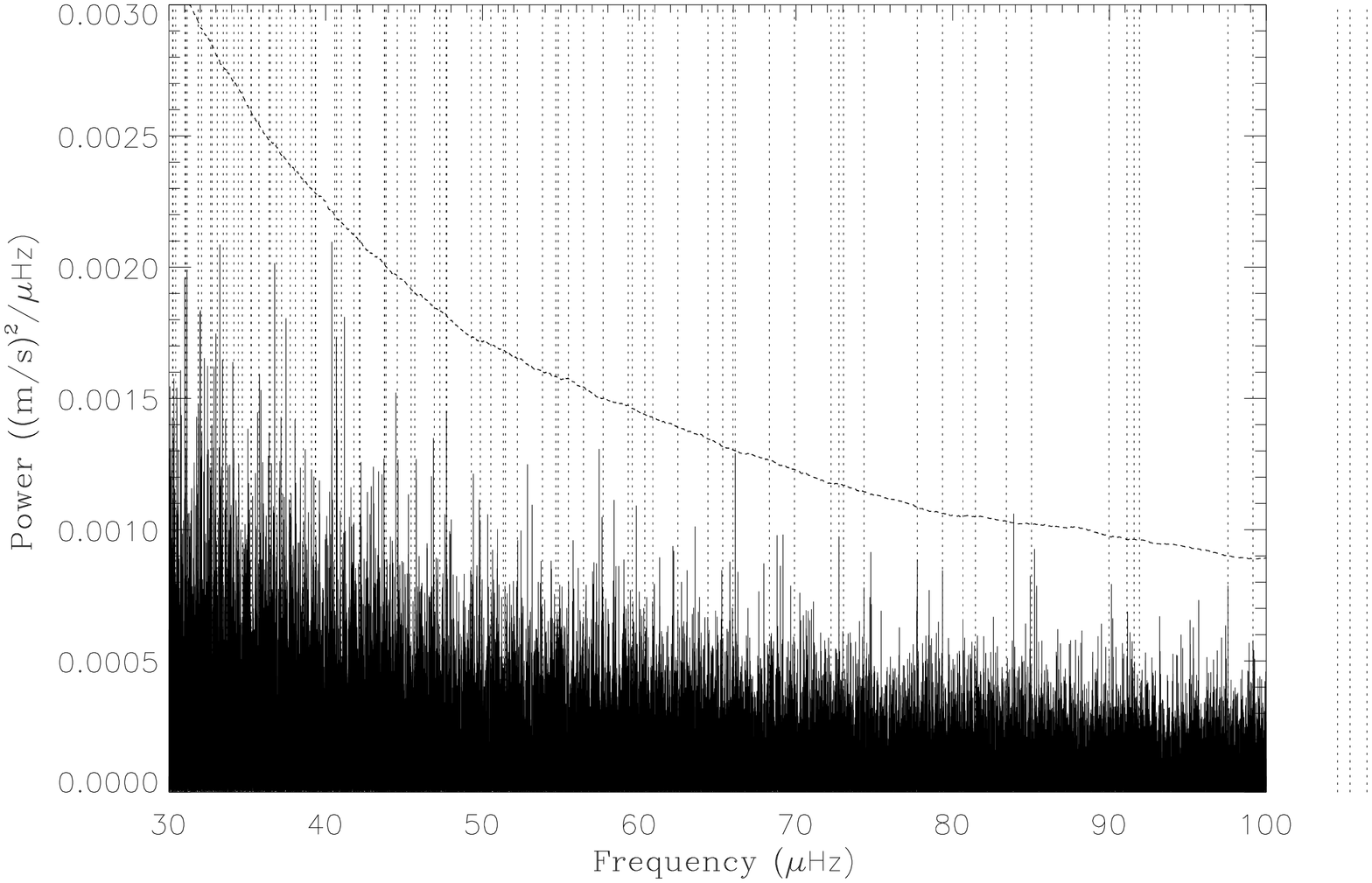}\\
\includegraphics[height=9cm,width=6cm,angle=90]{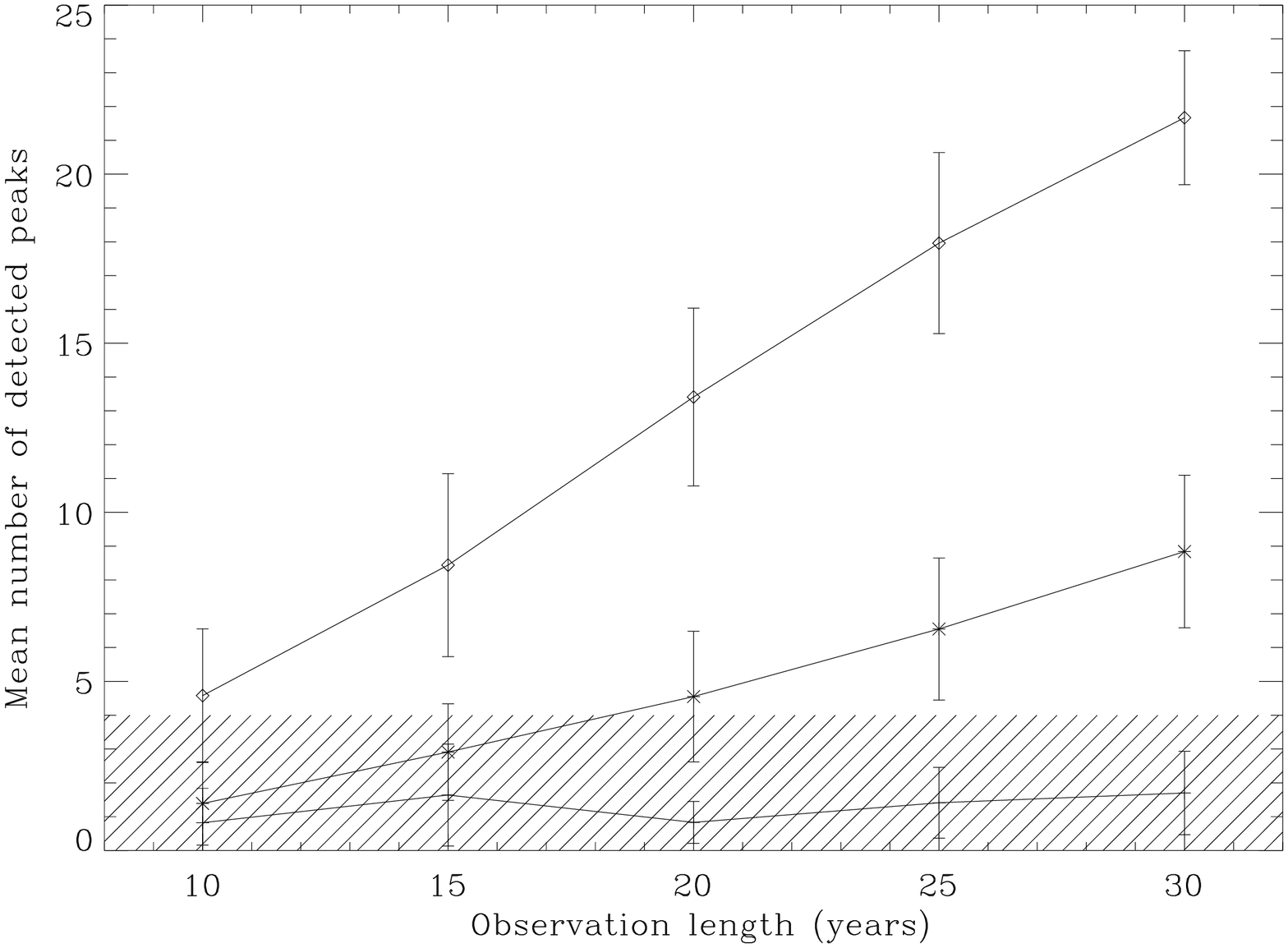}
\caption{
Simulated spectrum for an observation length of 30 years, in 
Case 1. The dashed line indicates the level of detection (see text and
Eq.~\ref{trucmuche}). The vertical lines indicate the frequencies of the simulated
modes. Here, only one mode is above the detection level.
{\bf Bottom:} 
Average number and standard deviation (from the 100 realizations) of
modes detected in simulations versus the length of observation for the three
cases (see text). Above the hashed region (less than four peaks detected), one can
consider the detection to be unambiguous. The upper curve corresponds to the 
case 3 ($A_{\rm max} = 2 A$), the middle one to case 2 ($A_{\rm max} = 1.5 A$), and 
the lower one to case 1 ($A_{\rm max} = A$).}
\label{plot_fred}
\end{figure} 

However, this approach has to be repeated for each mode (with its own proper noise level) 
to have a
global view of detection possibilities. To do so, we used simulations.
Again relying on the GOLF data to estimate the noise spectrum, we simulated
synthetic data including noise and g modes with the apparent velocities as
above (and with random phases). Several durations of observation were simulated,
from 10 to 30 years. A hundred simulations were performed in each case. The
noise level is estimated locally and so is the
detection level, following \eq{trucmuche}, on the frequency range
$[30\,\mu$Hz,$100\, \mu$Hz]. Thus, with a confidence level of 90\% and with 
7 independent subsets of 10$\mu$Hz, noise is expected to show no peak above the
global detection level with a probability of 48\%, and to show 1 peak above the
global detection level with a probability of 32\% (and even 2 peaks in
12\% of the realizations).
Table~\ref{machin} lists the average (over 100 simulations) number of
peaks detected above the detection level for different observation durations.
These simulations were performed using amplitudes $A_{max}$ assuming three different cases: 

$\bullet$ Case 1: we assumed  for $A_{max}$ the
apparent surface velocity amplitudes calculated  above, $A$.

Due to uncertainties in the theoretical modeling (as discussed in Sect.6), we also assume:

$\bullet$ Case 2: that  amplitudes  are 
larger than the amplitudes estimated above  
by 50\%  \emph{i.e.} $A_{max}= 1.5 ~A$

$\bullet$ Case 3:  that  amplitudes are 
larger than the amplitudes estimated above  
by  a factor 2 \emph{i.e.}  $A_{max}= 2 ~A$

\begin{table}
\begin{center}
\begin{tabular}{|c|c|c|c|c|c|}
\hline   $A_{max}$&10 years & 15 years & 20 years & 25 years & 30 years \\
\hline $A$  & 0.8 & 1.6 & 0.8 & 1.4 & 1.7 \\
\hline $1.5A$ & 1.4 & 2.9 & 4.5 & 6.5 & 8.8 \\
\hline $2A$  & 4.6 & 8.5 & 13.4 & 20.0 & 21.7 \\
\hline
\end{tabular}
\caption{\label{machin} Number of peaks above the detection level in the simulated power spectra
versus the duration of observation in three cases. In the simulated signal, the modes are given an amplitude $A_{max}$.
The 3 cases respectively correspond to  $A_{max}$ 
being the apparent amplitudes $A$ readily stemmed  from our calculation, $A_{max}=1.5A$ and  $A_{max}=2A$.
The last two cases take into account that uncertainties in the modeling globally 
 tend to underestimate   the amplitudes as discussed in Sect.6. 
}
\end{center}
\end{table}

Cases 1 and  3 are the two limits of this exercise. The number of
detected peaks in Case 3 shows that the predicted amplitudes cannot be
overestimated by a factor of two, because in this case, the solar g modes would have already 
been detected without doubt. Case 1 sets a lower limit, because in this case,
even with longer (30 years) observation, g modes would not be detected.  Case
2 shows that if 
real solar amplitudes are just a few tens of percent higher than the present estimations, 
then $g$ modes could be detected no doubt after say 15 to 20 years of observation 
(to be compared to the present status of observation: 12 years).  
The results are summarized in Figs.~\ref{plot_fred}. 

We must stress that, apart from visibility effects and height of line formation, 
we took  no other instrumental  effects on the apparent amplitude determination into account,  
because they depend on the instrument. The impact is probably a decrease in the measured amplitudes 
compared to the apparent amplitudes as computed here. This does not change the above conclusion for Case 1.  
We expect that the instrumental uncertainty is less than the theoretical uncertainties   discussed in Sect. 6 below,  
which led to Case 2 and 3. 

\section{Discussion}
\label{discussion}

In Sect.~5 above, we explained why estimates of  $g$-mode amplitudes obtained by previous authors differ from each other
by orders of magnitude \citep{CD2002}.  
We propose an improved modeling based on the input of 3D numerical simulations 
and on a formalism that had successfully reproduced the observations for $p$ modes \citep{Belkacem06b}.
Nevertheless,   several approximations  remain, and they lead to uncertainties that can reach a factor two 
in the estimation of $g$-mode apparent velocities (overestimation). 
We next discuss the most important ones.

\subsection{Equilibrium model: description of convection}

Convection is implemented  in our equilibrium models according to the classical  B\"{o}hm-Vitense mixing-length (MLT) formalism \citep[see][for details]{Samadi06}.

\subsubsection{ Convective velocities }

Values of the MLT convective velocity, $u$, are by far the most important contributions to mode amplitude uncertainties 
because the mode surface velocities depend on $u^3$ . 
First, we verified that a non-local description of turbulence does not modify the convective velocities by more than a few per cent except near the uppermost part of the convection zone, which  does not play any role here. Second,  
we compared the rms velocities from the 3D numerical simulation with  MLT velocities to estimate of the uncertainties.  
The MLT underestimates the velocity, relative to the more realistic numerical simulation (far from the boundaries). Indeed, it comes from the negative kinetic energy flux that results in a larger enthalpy flux in order to carry the solar flux to the surface. A direct consequence is that in 3-D simulations the velocities are higher than the ones computed by MLT by a factor of about $50 \, \%$. This may in turn result in a possible underestimation of the amplitudes of the modes by a factor 2, when, as here, MLT is used to estimate the velocities.

\subsubsection{Anisotropy}

 The  value for the velocity anisotropy, which is the ratio between the square of the vertical velocity to the square of the rms velocity parameter, $\Phi$,  is derived from the MLT: its value is $2$. However, 
this is not fully consistent  since we assume, in the excitation model, isotropic turbulence (i.e. $\Phi=3$). 
Nevertheless, increasing the value of $\Phi$ from two to three  results in an increase of only $15 \, \%$ 
in the mode surface velocities. This  is lower than the uncertainties coming from $\chi_k$ (see Sect.~\ref{chi_k}).

\subsubsection{ Turbulent pressure }

Our solar equilibrium model  does not include  turbulent pressure. However, unlike  $p$ modes, 
low-frequency (high radial order) gravity modes, i.e. those considered in this work, are only slightly affected by turbulent pressure. 
The reason is that such modes are excited in the deepest layers of the convection zone, i.e. between $r=0.7 \, R_\odot$ and 
$r=0.9 \, R_\odot$ where turbulent pressure has little influence on the equilibrium structure since the ratio of the turbulent pressure to the gas pressure increases
 with the radius.

\subsection{Stochastic excitation: the  role of the eddy-time correlation function}
\label{chi_k}

A Gaussian function is commonly used to describe the frequency dependence of the turbulent kinetic energy spectrum, 
$\chi_k$ , \citep[e.g., ][]{Samadi00I,Chaplin05}. However,  \cite{Samadi02II}  
show that, for $p$ modes, a Lorentzian function represents the results obtained using
 3D numerical simulations better. Furthermore,  the latter function yields a theoretical modeling in accordance with observations, 
 while using a Gaussian function fails \citep{Samadi02I}. This led us to investigate $\chi_k (\omega)$ for $g$ modes. 
 We find that different  choices of the functional form for  $\chi_k (\omega)$  result in order of magnitude differences for the mode amplitudes.

 Uncertainties inherent in the eddy-time correlation function are related to the value of the   $\lambda$  parameter (Sect.~\ref{kin}) and to the contribution of low frequency components in the 3D simulation. 
As a rough estimate, decreasing  $\lambda$  from 3 to 2 leads to an increase of $20\,\%$ for the surface velocity.  
Figure~\ref{chik_simu_vs_analytique} shows that   low-frequency components in the turbulent kinetic energy spectrum are 
better-fitted using a Gaussian function. 
However, the source of such low-frequency components remains unclear, because they can originate from rotation; in particular,
it is not clear whether they must be taken into account when
estimating  the  mode 
excitation rates. By removing those contributions, the resulting surface velocities decrease by around $25\,\%$.

\subsection{Mode damping: the convection-pulsation coupling}

Last but not least, modeling damping rates of damped, stochastically excited modes remains one of the most challenging issues. 
 The strong coupling between convection and oscillation in solar-like stars makes the problem  difficult and still unsolved, 
 since all approaches developed so far failed to reproduce the solar damping rates without the use of unconstrained free parameters (e.g., Dupret et al. 2005, Houdek 2006). Such descriptions fail to correctly describe the interaction between convection and oscillations when both are strongly coupled, \emph{i.e.} when the characteristic times associated with the convective motions are the same order of magnitude as the oscillation periods.
This explains why we do not use an extrapolation based on a fit of $p$ mode damping rates, but instead consider a frequency domain in which the damping  is dominated by radiative contributions. A reliable computation  of  the damping rates at higher frequencies, beyond this paper's scope,  would 
require a sophisticated analytical or semi-analytical  theory of the convection-oscillation interaction, which will not be limited 
to the first order in the convective fluctuations and which will take the contribution of different spatial scales into account.

\section{Conclusions}

We performed a theoretical computation of the surface oscillation velocities of asymptotic gravity modes. 
This calculation requires knowing  excitation rates, which were obtained  as described in \cite{Belkacem08} 
with input from 3D numerical simulations  of the solar convective zone \citep{Miesh08}. 
Damping rates, $\eta$ , are also needed.  As mentioned in Sect.~\ref{discussion}, we restricted our investigation
to the frequency domain for which $\eta$ is dominated by radiative contributions (\emph{i.e.} $\nu \in [20;110] \mu$Hz).
For higher frequencies, the coupling between convection and oscillation becomes dominant, making the
theoretical predictions doubtful.
For asymptotic $g$-modes, we find that damping rates are dominated by the modulation 
of the radial component  of the radiative flux by the oscillation. 
In particular for the $\ell=1$ mode near $\nu \approx 60 \, \mu$Hz, $\eta$ is around  $10^{-7} \, \mu$Hz, 
then the mode life time is $\approx 3. 10^{5}$yrs. Maximum velocity amplitude at the photosphere arises  
for this same mode and  is found at the level of  3 mm\,s$^{-1}$ (see Fig.~\ref{vitesses}). 
Modes with higher values of the angular degree $\ell$ present smaller amplitudes since the damping is 
proportional to $\ell^2$. 

Amplitudes found in the present work are orders of magnitude larger 
than those from  previous works, which themselves showed a large dispersion
between their respective results. In one of these previous works,  
the estimation was based on an equipartition principle 
derived from the work of \cite{GK77a,GK77} and designed for $p$ modes. Its use for asymptotic $g$ modes is not adapted 
as the damping rates  of these modes  are not dominated by turbulent viscosity. 
\cite{Kumar96} have carried  another investigation of $g$ mode amplitudes, and its calculation is rather close to 
our modeling. Most of the quantitative disagreement  with our result lies in the use of 
a different  eddy-time correlation function. \cite{Kumar96} assumed a Gaussian function as is  commonly used. Our
choice relies on results from 3D simulations  and is closer to a Lorenztian function.

Taking  visibility factors, as well as the limb-darkening, into account  
we finally found that the maximum of apparent surface velocities of asymptotic
 $g$-modes is $\approx 3$ mm\,s$^{-1}$ for $\ell=1$ at $\nu\approx 60\,\mu$Hz an $\ell=2$ at 
$\nu\approx 100\,\mu$Hz. 
Due to uncertainties in the theoretical modeling, 
amplitudes at maximum, \emph{i.e.} for $\ell=1$ at 60 $\mu$Hz, can range from 3 to 6 mm\,s$^{-1}$. 
By performing numerical simulations of power spectra, it is shown that, with amplitudes of  6 mm\,s$^{-1}$,
 the modes would have been already detected by the GOLF instrument, while in the case of 
 an amplitude of $3$ mm\,s$^{-1}$ the $g$ modes would remain undetected even with 
 30 years of observations. The theoretical amplitudes found in this work are then close to the actual observational limit. 
 When detected, the amplitude detection threshold of these modes will, for instance, establish  a strict upper limit to the convective velocities in the Sun.
 
\begin{acknowledgements}
We are indebted to J.~Leibacher for his careful reading of the manuscript and his helpful remarks.
We also thank J.P~Zahn for its encouragements.
\end{acknowledgements}

\newpage 

\appendix
\section{Energy equation near the center}
\label{enregy_eq}

For the full non-adiabatic computation of $g$-mode damping rates, 
much care must be given to the solution of the energy equation near the center of the Sun 
 for the modes of angular degree $\ell=1$. 
We give in Eqs.~(\ref{E}) and (\ref{diff}) the perturbed energy and transfer equations 
in a purely radiative zone: 

{\setlength\arraycolsep{2pt}
\begin{eqnarray} \label{E}
\mathrm{i}\,\omega_0\,T\,\delta S &=& 
-\frac{\textrm{d}\,\delta L}{\textrm{d} m}
+\epsilon\left(\frac{\delta\epsilon}{\epsilon}
+\frac{\delta\rho}{\rho}+\frac{1}{r^2}\frac{\textrm{d}\,(r^2\xi_r)}
{\textrm{d}r}\right)\nonumber \\
 &+&\ell(\ell+1)\frac{L}{4\pi\rho r^3}\left(\frac{\delta T}
{r\,\mathrm{d}T / \mathrm{d}r} - \frac{\xi_r}{r} \right)\,, 
\end{eqnarray}} 
\begin{equation} \label{diff}
\frac{\delta L}{L} = 2\,\frac{\xi_r}{r} + 
3\, \frac{\delta T}{T}
 - \frac{\delta\kappa}{\kappa} - \frac{\delta\rho}{\rho}   
+\frac{1}{\textrm{d}T/\textrm{d}r} \frac{\textrm{d} \delta T}{\textrm{d} r}
- \frac{\textrm{d}\xi_r}{\textrm{d} r} \,.
\end{equation}

The radial (first term of Eq.~(\ref{E})) and transverse parts (last term of  Eq.~(\ref{E})) 
of the perturbed flux divergence are both singular at the center. But this singularity is lifted 
when the two terms are joined and an appropriate change of variables is carried out: 
\begin{eqnarray} 
 \sigma &=& {\omega_0 \over \sqrt{GM/R^3}} \cr
{\xi_r \over r} &=&  \zeta  ~ x^{\ell-2}\cr
{\delta s \over c_v}&=& \eta  ~x^\ell  ~~~~;~~~~~ {\delta T \over T} = \vartheta ~x^\ell ~~~;~~~~ 
{\delta \rho \over \rho}  = \gamma  ~x^\ell  \cr 
{\delta\epsilon \over \epsilon} &=&   \delta \epsilon_x ~ x^\ell\cr
 k&=& (G M/R^3)^{-1/2}\, {L(r)\over 4\pi\rho r^3 c_v }\cr 
 \epsilon_1 &=&({4\pi \rho r^3\over 3} {\epsilon \over L(r)}-1)\,{3 \over x^2}\cr  
T_1 &=& {x \over \mathrm{d} \ln T/\mathrm{d}x} \cr 
T_2 & =&{x^2 \over L}\: {\mathrm{d}\over {\mathrm{d} x}}
 \left( {L\over x^2} {1\over \mathrm{d}\ln T/\mathrm{d}x}\right) \cr
 x &=& \frac{r}{R} \, .
 \end{eqnarray}
 
All of these variables and quantities are regular at the solar center,
 where the perturbed energy equation takes the form
\begin{eqnarray} \label{Ec}
\frac{i\,\sigma\,\eta}{k} &=& 3\,\delta\epsilon_x\:+\:2\,\gamma 
\nonumber\\
&-&(\ell+3)\,\left(\,(4\,-\,\kappa_T)\,\vartheta \:-\: 
(1\,+\,\kappa_{\rho})\,\gamma\,\right) \nonumber \\
&-& \ell\,T_2\;\vartheta 
\:-\: (2 \ell+3)\:T_1\:\frac{{\mathrm{d}}^2\vartheta}{\mathrm{d} x^2} \nonumber \\
&+& 2(\ell-1)\epsilon_1 \,\zeta
\:+\:(2\ell+3)
\:\frac{{\mathrm{d}}^2\zeta}{\mathrm{d} x^2}\;.
\end{eqnarray}

For a precise solution of the non-adiabatic problem by a finite difference method, it is crucial to use a discrete 
scheme that tends continuously towards Eq.~(\ref{Ec}) at the center. If not, the eigenfunctions diverge towards 
the center; in the particular case of the solar $g$ modes, this can lead to an overestimate of the damping 
rates by a factor of about 2.   

\section{Computation of the kinetic energy spectrum from the ASH code}
\label{ASH}

The ASH code solves the hydrodynamical equations in spherical coordinates $(r,\theta,\phi)$.
Each component of the velocity  is decomposed in terms of 
spherical harmonics as
\eqn{
 \label{Vp}
V_p (t,r,\theta,\phi) = \sum_{{\cal \it l},m} \, V_{{\cal \it l},m,p}(t,r) \, Y_{{\cal \it l},m} (\theta,\phi)
}
where $p=r,\theta,\phi$. The spherical harmonic $ Y_{{\cal \it l},m}(\theta,\phi)$
 is defined as
\begin{equation}
Y_{\cal \it l}^m(\theta,\phi) \equiv N_{{\cal \it l},m} \, P_{\cal \it l}^m(\cos \theta) \, e^{i  m \phi}
\label{ylm}
\end{equation}
where $P_{\cal \it l}^m$ is the associated Legendre function, and the normalization
 constant $N_{{\cal \it l},m}$  
\begin{equation}
N_{{\cal \it l},m}= \sqrt{{2{\cal \it l}+1\over 4\pi}}~  \sqrt{{({\cal \it l}-m)!\over ({\cal \it l}+m)!}} 
\label{Nlm}
\end{equation}
is chosen such that 
\eqn{
\int d\Omega \,  Y_{{\cal \it l},m} (\theta,\phi) \, Y_{{\cal \it l}^\prime,m^\prime}
(\theta,\phi) = \delta_{{\cal \it l},{\cal \it l}^\prime} \,
\delta_{m,m^\prime}
\label{othogonalite}
}
where $d\Omega= \sin\theta d\theta d\phi$.

The   kinetic energy spectrum that is   averaged over time and the  solid
angle is defined following \citet{Samadi02I} as
\eqn{
E(\ell,r)   \equiv  {1 \over 2}  \, 
%\int {d\Omega \over {4 \pi}} \, 
\sum_{m,p} \, \left \langle \left (  V_{{\cal \it l},m,p} 
- \left \langle  V_{{\cal \it l},m,p} \right
\rangle  \right )^2 \right \rangle
\label{elta}
}
where $\left \langle (.) \right \rangle$ refers to time average.
As in \citet{Samadi02I}, density does not enter into the
definition of the kinetic energy spectrum. Indeed, the   
\citet{Samadi00I}' formalism assumes a homogeneous  turbulence.
This assumption is justified when the turbulent Mach number is low.
This is the case in most parts of the convective zone except at the
top of convective region.

The mean kinetic energy spectrum, $ E({\cal \it l},r)$, verifies the
relation
\eqn{
\sum_{{\cal \it l}} E({\cal \it l},r) =  {1 \over 2} \, u^2(r)
}
where $u(r)$ is the root mean square velocity at the radius $r$.

Following \citet{Samadi02II}, we also define   a kinetic energy spectrum as
a function of frequency ($\nu$) and  averaged
over the solid angle, $E({\cal \it l},\nu,r)$ such that 
\eqn{
\sum_{{\cal \it l}} ~E({\cal \it l},\nu,r)   \equiv   { 1 \over
    2} \,  
\int {d\Omega \over {4 \pi}} \, 
\sum_{p} \, \left \|  \hat V_{p} (\nu,r,\theta,\phi)
\right  \|^2 
\label{elnu}
}
where $\hat V_p(\nu,r,\theta,\phi)$ is the time Fourier transform of
$V_p(t,r,\theta,\phi) - \left \langle V_p \right \rangle$. 
Using Eqs.(\ref{Vp}) and (\ref{othogonalite}), Eq.~(\ref{elnu})
yields:
\eqn{
E({\cal \it l},\nu,r) =  {1 \over 2} \,
\sum_{m,p} \, \left  \|   \hat V_{{\cal \it l},m,p} (\nu,r)
\right \| ^2 
\label{elnu:2}
}
where $\hat V_{{\cal \it l},m,p}(\nu,r) $ is the time Fourier transform of
$V_{{\cal \it l},m,p}(t,r) -  \left \langle V_{{\cal \it l},m,p} \right \rangle$.
As in \citet{Samadi02II}, we decompose $E({\cal \it l},\nu,r)$ as 
\eqn{
E({\cal \it l},\nu,r) = E({\cal \it l},r) \; \chi_{\cal \it l}(\nu,r)
\label{elnu:3}
}
where the function $\chi_{\cal \it l}(\nu,r)$ satisfies the normalization condition
\eqn{
\int_{-\infty}^{+\infty}d\nu \, \chi_{\cal \it l}(\nu,r) = 1 \, .
}
According to the Parseval-Plancherel relation, one has 
\eqn{
\sum_{\cal \it l} \, \int_{-\infty}^{+\infty}d\nu \, E({\cal \it
  l},\nu,r) = \sum_{{\cal \it l}} E({\cal \it l},r) = {1 \over 2 } \, u^2(r) \, ,
}
We consider a short time series of duration $\approx$ 4.68 days with a sampling time
of 800 seconds. Accordingly the Nyquist frequency is $\approx$ 1~mHz and the frequency
resolution reachs $\approx$ 2.5~$\mu$Hz. In addition, we use a longtime series of 
duration $\approx 45.83$  days with a sampling time of $4 \,10^4$ seconds that permits us to 
get $\chi_k$ at very low frequencies.
In practice, $E({\cal \it l})$ is derived from \eq{elta} and is directly implemented into \eq{puissance}, while  
$\chi_k(\nu)$ inferred from the simulation is computed using Eqs.~(\ref{elnu:3}) and (\ref{elta}).

By using $E({\cal \it l})$ from the numerical simulation, we assume a planparallel 
approximation ($E(k)\, \textrm{d}k=E({\cal \it l}) \, \textrm{d}{\cal \it l}$) since the maximum of the kinetic energy spectrum 
occurs on scales ranging between ${\cal \it l} \approx 20$ and ${\cal \it l} \approx 40$.

\section{Visibility factors} 
\label{visibilite}

Visibility factors have been computed first by \cite{Dziembowski77}. 
\cite{Berthomieu90} studied the case of g modes  which, for convenience, we recall  below in our own notation.
We denote the spherical coordinate system in the observer's frame  by $(r,\theta,\phi)$ 
where $r=0$ corresponds to the center of the star and the $\theta=0$ axis coincides with the observer's direction. 
At a  surface point   $(r,\theta,\phi)$, the unit vector directed toward the observer is
   $\vec n = \cos \theta \, \vec e_r  -  \sin \theta \, \vec e_\theta$. 
The apparent surface velocity is obtained as
\begin{eqnarray}
V_{app} (r,t)=  \frac{\int  h(\mu) \, \left( \vec v(\vec r,t) \cdot \vec n \right)\textrm{d} \, \Omega } 
{\int  \,  h(\mu) ~ \textrm{d}\Omega} \, ,
\label{Vobs}
\end{eqnarray}
where $\vec v(\vec r,t)$ is the intrinsic mode velocity 
 and $h(\mu)$ the limb-darkening function, which  is normalized such that:
\eqn{
 \int_0^1 \mu~ h(\mu) d\mu  = 1 \, .
}
To first order in linearized quantities
 in Eq.~(\ref{Vobs}),  the effect of the distorted surface is neglected, 
and $\textrm{d}\Omega=  R^2 \sin \theta \textrm{d}\theta \textrm{d}\phi$ is the solid angle around the direction 
 of the observer $\vec n$ with $R$ the stellar radius.

For slow rotation, the oscillation velocity can be described in a pulsation frame with a single spherical harmonic. 
The coordinate system  $(r, \Theta, \Phi)$ in the pulsation frame is  chosen 
such that  the pulsation polar axis coincides   with the rotation polar axis.
The  velocity  vector at a level $r$ in the atmosphere of the star for a mode 
with given $\ell,m$  and pulsation frequency $\omega_0$  can then be written with no loss of generality as
\begin{align}
\label{vitesse_annexe}
\vec v(\vec r,t) = {1\over 2} ~ a(t) ~\omega_0 ~\vec \xi(\vec r) ~ e^{i\omega_0 t}+c.c.
\end{align}
where $c.c.$ means complex conjugate and with the displacement eigenvector  defined as 
\begin{align}
\label{depl}
\vec \xi(\vec r) = \xi_r(r) ~Y_\ell^{m} (\Theta, \Phi) ~\vec e_r+  \xi_h(r) ~ \vec \nabla_H Y_\ell^m (\Theta, \Phi)
\end{align}
with
\begin{align}
\vec \nabla_H=(0,{\partial\over \partial \Theta}, {1\over \sin \Theta}{\partial\over \partial \Phi}) \, .
\end{align}

The dimensionless complex velocity amplitude $a_v(t)$ is assumed to be a slowly varying function 
of time for a damped stochastically excited mode \citep{Samadi00I,Samadi02I,Belkacem08}. The 
theoretical expression is given by 
\begin{align}
<|a(t)|^2> = {P\over  \eta I \omega_0^2}
\label{amp} 
\end{align}
where the power $P$ is defined in \eq{puissance}, $I$ is the mode inertia, $\eta$ the damping rate and $<>$ 
represents a statistical average, or equivalently here a time average.

To obtain the apparent velocity from  \eq{Vobs} using Eqs.~(\ref{vitesse_annexe}) and 
(\ref{depl}), 
 one must compute the scalar product: $\vec \xi(\vec r) \cdot \vec n $. 
\begin{align}
\label{xi}
 \vec \xi(\vec r) \cdot \vec n = \xi_r(r) ~Y_\ell^m (\Theta, \Phi) ~(\vec e_r \cdot \vec n)+  
 \xi_h(r) ~(\nabla_H Y_\ell^m \cdot \vec n) \, .
 \end{align} 
A change in coordinate system shows that $\vec e_r \cdot \vec n = \cos \theta$ and 
$$ \nabla_H Y_\ell^m(\Theta, \Phi) \cdot \vec n = -\sin \theta {\partial Y_\ell^m(\Theta, \Phi)\over \partial \theta}$$  
We use the  spherical harmonics as  defined in \eq{ylm} and the following property 
\begin{align}
\label{harm_01}
P_\ell^m (\cos \Theta) e^{im\Phi}= \sum_{m'=-\ell}^{\ell} ~q^\ell_{m,m'}(\Theta_0,\Phi_0) P_\ell^{m'} (\cos \theta) ~e^{im'\phi} \, ,
\end{align}  
which for convenience, we use under the form 
\begin{align}
\label{harm_01}
Y_\ell^m (\Theta, \Phi) = N_{\ell,m}  \sum_{m'=-\ell}^{\ell} ~q^\ell_{m,m'}(\Theta_0,\Phi_0)  ~P_\ell^{m'} (\cos \theta)  ~e^{im'\phi} 
\end{align}  
where $N_{\ell,m}$ is defined in \eq{Nlm}, and $(\Theta_0,\Phi_0)$  are the coordinates of the line-of-sight direction 
in the pulsation frame. The scalar product \eq{xi} becomes 
\begin{eqnarray}
\label{xi2}
\vec \xi(\vec r) \cdot \vec n &=& N_{\ell,m}   \sum_{m'=-\ell}^{\ell} ~q^\ell_{m,m'}(\Theta_0,\Phi_0)  ~
~e^{im'\phi} \nonumber\\
&\times& \left( \right.\xi_r(r) ~P_\ell^{m'}  ~\cos \theta -  \xi_h(r) ~\sin \theta { \textrm{d}P_\ell^{m'}\over \textrm{d}\theta}
\left.\right) \, .
 \end{eqnarray} 

As emphasized by \cite{Dziembowski77}, only the $q^\ell_{m,0}$ coefficients survive the
 $\phi$ integration in \eq{Vobs}. Its expression is
\begin{align}
q^\ell_{m,0}(\Theta_0,\Phi_0)  =  P_\ell^{m}(\cos \Theta_0)~ e^{im\Phi_0} \, .
\end{align}
The angle $\Theta_0$ between the observer and the rotation axis is often denoted  $i$. 
Integration over the solid angle leads to :
\begin{eqnarray}
&&\int  h(\mu) (\vec \xi(\vec r) \cdot \vec n ) d\Omega =   ~Y_\ell^m(\Theta_0,\Phi_0)  \times \nonumber \\
 & & \Bigl(\Bigr.\xi_r(r) ~ \int_0^1 \mu^2  ~h(\mu)  ~P_\ell(\mu)  d\mu  + \nonumber \\
  & & \xi_h(r) \int_0^1 \mu ~ h(\mu) ~ (1-\mu^2)~ {dP_\ell (\mu) \over d \mu}
d\mu \Bigl. \Bigr) 
\end{eqnarray}
Finally,  using  properties of spherical harmonics, one obtains 
\begin{equation}
{\int  h(\mu) (\vec \xi(\vec r) \cdot \vec n ) d\Omega  \over \int  h(\mu)d\Omega  }=  
Y_\ell^m(\Theta_0, \Phi_0) ~  \left(\xi_r(r) ~ u_\ell + \xi_h(r) ~w_\ell\right)
\label{scalp}
\end{equation}  
where we have defined
\begin{eqnarray}
u_\ell &=& \int_{0}^{1} \textrm{d} \mu \,  \mu^2  \tilde h(\mu)  \, P_\ell (\mu) \\
w_l &=&  \ell \int_{0}^{1} \textrm{d} \mu \, \mu  \; \tilde  h(\mu) \left( P_{\ell-1} - \mu P_{\ell} \right)  
\end{eqnarray}
with 
\begin{equation}
\tilde h(\mu)={h(\mu) \over \int_0^1  h(\mu)d\mu} \, .
\end{equation}

Collecting \eq{vitesse_annexe}  and \eq{scalp}, the apparent velocity is then given by 
%\eq{harm_01} 
%and \eq{harm_02} leads to
\begin{align}
V_{app} (r,t) &=  {1\over 2} ~ a(t) ~\omega_0 ~ N_{\ell,m}~ P_\ell({\cos \Theta_0}) \\
&\times \left(\xi_r(r) u_l+ \xi_h(r) w_l\right) e^{i(\omega_0 t+m\phi_0)}  +c.c. \, .
\end{align}
We assume a quadratic limb-darkening law of the form 
\eqn{
h(\mu) = 1 + c_1 ~X^2 + c_2 ~X^2 + c_3 ~X^3
}
where $X=1-\mu$, $c_{i=\{1,2,3\}}$ are the associated limb-darkening coefficients, 
which respective values are $-0.466$, $-0.06$ and $-0.29$ for the NaD1 spectral line, 
as derived by \cite{Ulrich00}.  We find that our conclusion  depends neither on the adopted limb-darkening
law nor  on the limb-darkening coefficients, results in accordance 
with \cite{Berthomieu90}.

Using \eq{amp}, the rms velocity is obtained as: 
\begin{align}
(<|V_{app} (r,t)|^2>)^{1/2}&=   \left({P\over 2 \eta I \omega_0^2}\right)^{1/2} ~\omega_0 ~ N_{\ell,m}~|P_\ell({\cos \Theta_0})|\\ \nonumber
&\times |\xi_r(r) ~u_l+ \xi_h(r) ~w_l| 
\end{align}
which we finally rewrite as 
\begin{align}
(<|V_{app} (r,t)|^2>)^{1/2}&=    \left({P\over 2 \eta I \omega_0^2}\right)^{1/2}  \\ \nonumber 
&\times |v_r(r) ~\alpha_{\ell}^{m}+ v_h(r) ~\beta_{\ell}^{m}| \, ,
\end{align}
where we have defined 
\begin{eqnarray}
\alpha_{\ell}^{m} &=& N_{\ell,m}~ P_\ell({\cos \Theta_0})~ u_l   \\
\beta_{\ell}^{m} &=&  N_{\ell,m}~~P_\ell({\cos \Theta_0}) ~w_l \, ,
\end{eqnarray}
and
\begin{equation}
v_r(r)= \omega_0 ~ \xi_r(r) ~~;~~ v_h(r)=\omega_0 ~ \xi_h(r) \, .
\end{equation}

\end{document}